\documentclass[prd,twocolumn,preprintnumbers,nofootinbib]{revtex4}
\usepackage{graphicx}
\usepackage{dcolumn}
\usepackage{bm}
\usepackage{color}
\usepackage{amssymb}
\usepackage{amsmath}
\usepackage{url}

\def\kms{\,{\rm km}\,{\rm s}^{-1}}

\def\mpch{\,{h {\rm Mpc}^{-1}}}
\def\hubble{\, \kms \,{\rm Mpc}^{-1}}
\def \der{{\rm d}}

\begin{document}

\title{Dark energy imprints on the kinematic Sunyaev-Zel'dovich signal}

\author{Yin-Zhe Ma}
\email{mayinzhe@phas.ubc.ca} \affiliation{Department of Physics
and Astronomy, University of British Columbia, Vancouver, V6T 1Z1,
BC Canada.} \affiliation{Canadian Institute for Theoretical
Astrophysics, Toronto, Canada.}

\author{Gong-Bo Zhao}
\email{gongbo@icosmology.info}
\affiliation{National Astronomy
Observatories, Chinese Academy of Science, A20 Datun Road,
Chaoyang District, Beijing, China}
\affiliation{Institute of Cosmology and Gravitation, University of Portsmouth, Dennis Sciama
Building, Portsmouth, PO1 3FX, UK}

\begin{abstract}
We investigate the imprint of dark energy on the kinetic
Sunyaev-Zel'dovich (kSZ) angular power spectrum on scales of
$\ell=1000$ to $10000$, and find that the kSZ signal is sensitive
to the dark energy parameter. For example, varying the constant
$w$ by 20\% around $w=-1$ results in a $\gtrsim10\%$ change on the
kSZ spectrum; changing the dark energy dynamics parametrized by
$w_a$ by $\pm0.5$, a 30\% change on the kSZ spectrum is expected.
We discuss the observational aspects and develop a fitting formula
for the kSZ power spectrum. Finally, we discuss how the precise
modeling of the post-reionization signal would help the
constraints on patchy reionization signal, which is crucial for
measuring the duration of reionization.

\end{abstract}


\maketitle

\section{Introduction}
\label{sec:intro}

Dark energy (DE), the energy source that drives our Universe
accelerating, has remained a mystery since it was
discovered in 1998 \cite{Riess98,Perlmutter99}. 
The key feature of dark energy is encoded in its equation of state
(hereafter EoS) parameter $w$, which is the ratio of its pressure
to the energy density. The time dependence of EoS can
be used to classify a range of DE models. The accumulating
observational data, including observations of the cosmic microwave
background radiation (CMB) \cite{Hinshaw12,Planck16}, Type-Ia
supernovae (SN) data \cite{Conley11,Suzuki12} and baryon acoustic
oscillation (BAO) from galaxy surveys
\cite{Beutler11,Padmanabhan12,Anderson12,Blake12} have set up
strong constraints on the EoS of dark energy. Assuming the dark
energy EoS is a constant, then recent observation from
\textit{Wilkinson Microwave Anisotropy Probe} (\textit{WMAP})
gives the constraint $w=-1.073 \pm 0.180$ ($95\%$ confidence
level, \textit{WMAP}9+extra CMB data \footnote{This ``extra CMB
data'' refers to the band-power spectra data from 150GHz
South Pole Telescope (SPT) \cite{Keisler11} and 148GHz Atacama
Cosmology Telescope (ACT) \cite{Das11}.} +BAO+$H_{0}$,
\cite{Hinshaw12}), and observation from \textit{Planck} satellite
gives $w=-1.24^{+0.18}_{-0.19}$ ($95\%$ CL from
\textit{Planck}+$H_{0}$+\textit{WMAP} polarization data,
\cite{Planck16}). However, allowing time evolution of $w$, the
results of constraints become comparatively looser. For example, 
parameterizing dark energy EoS as $w(a)=w_{0}+w_{a}(1-a)$ then
the constraints from \textit{WMAP}9+extra CMB data +BAO+SN+$H_{0}$
is $w_{0}=-1.34 \pm 0.36$ and $w_{a}=0.85 \pm 0.94$ ($95\%$ CL,
see table~10 of \cite{Hinshaw12}), and from
\textit{Planck}+$H_{0}$+\textit{WMAP} polarization data it is
$w_{0}=-1.04^{+0.72}_{-0.69}$ and $w_{a}<1.32$ at ($95 \%$ CL).
Therefore, the data slightly favor the model with $w_{0}<-1$ and
$w_{a}>0$ while large uncertainties of parameters still exist in
the recent observational constraints.

In the spirit of exploring more phenomena associated with dark
energy, we would like to investigate how the dark energy affects
the growth of structure and clustering properties of galaxies. The
kinematic Sunyaev-Zel'dovich (hereafter kSZ, or kinetic SZ) effect
is one of the important phenomena that relates the galaxy's
peculiar motion with the temperature fluctuations of the CMB. The
effect can arise during two processes, i.e. consisting of the
``inhomogeneous patchy reionization'' and the post-reionization
signals.

In models of inhomogeneous reionization (or ``patchy
reionization''), where different regions of the Universe were
ionized at different times, the bulk motion of bubbles of free
electrons around the UV emitting sources may cause the temperature
anisotropy on the CMB
\cite{Gruzinov98,Fan06,Iliev06,Knox98,McQuinn05,Santos03,Zahn05,Zahn12}.
It has been demonstrated \cite{Zahn05,McQuinn05} that the
magnitude of the kSZ power from patchy reionization is related to
the duration of reionization. Hence, one can set a constraint on
the duration of reionization ($\Delta z_{\rm rei}$) once the
optical depth to reionization can be measured \cite{Zahn12}. After
reionization, the ``secondary anisotropy'' of CMB can also be
generated from the peculiar motion of galaxy clusters. Thus by
measuring the kSZ effect one can have a good handle on the
peculiar velocity of galaxies and therefore infer the growth rate
of large scale structure. The growth rate of the large scale
structure is affected by the dark energy EoS, because the dark
energy negative pressure can drive the accelerated expansion of
the Universe and therefore halt the growth of structure at late
times. Therefore, it is necessary to investigate the effect of
dark energy on growth of structure and the ``imprint'' of dark
energy on the kSZ effect. This research is particularly useful
since many ongoing CMB experiments, such as South Pole Telescope
(SPT and SPTPol \cite{Reichardt12}) and Atacama Cosmology
Telescope (ACT and ACTPol \cite{Dunkley11,Das11}) are going to
measure the kSZ effect to a high precision.

The effect of clustering can be reflected in three different
channels. First, the dark energy can freeze the growth of
structure at late times, the larger the density is, the earlier it
will take over the cosmic budget. Thus by counting the number of
galaxy clusters from SZ effect one can set up constraints on the
dark energy EoS \cite{Mak12}. Since the thermal SZ effect is
sensitive to the structure growth rate, another channel is to
measure the growth rate by cross-correlating the thermal SZ effect
with the galaxy clusters \cite{Hajian13}. Finally, due to the
change of the structures' growth rate, dark energy can
effectively change the power spectrum of kSZ effect. Thus by
computing the kSZ power spectrum, one can directly
measure the effect of dark energy from different $\ell$s of kSZ
power spectrum. Providing such an investigation on how much dark
energy effect on kSZ signal is the main aim of this paper. Such
detail modeling of post-reionization signal is particularly
meaningful as more precise CMB observations are measuring the
arcmin scale fluctuations. This is because once the astrophysics
of post-reionization era is known better, it is possible to
separate the post-reionization signal from the total signal, and
thus obtain a reliable constraint on patchy reionization signal
$\Delta z_{\rm rei}$. In addition, complicated simulation tool is
now developing to probe the physics of patchy reionization
\cite{Iliev06}.

This paper is organized as follows. In
Section~\ref{sec:kSZ_model}. we provide an overview of the kSZ
effect, and describe our model of the kSZ power spectrum, and
discuss the baryon gaseous pressure and patchy reionization effect
that may affect the shape and amplitude of power spectrum. In
Section~\ref{sec:dark_energy}, we explore different phenomena of
dark energy, by investigating how the different EoS functions
$w(z)$ can affect the the structure growth function and power
spectrum. Then in Section~\ref{sec:kSZ_DE}, we put together the
time evolution of dark energy and kSZ models and investigate how
the evolution of dark energy affect the 3D power spectrum of kSZ
and therefore affects its angular power spectrum. We then compare
our theoretical calculation with the current observational
constraints on kSZ, and discuss its relation to patchy
reionization signal. Our conclusion is presented in the last
section.

Except when referring to specific models with particular
parameters, throughout the paper we adopt a spatially
flat, $\Lambda$CDM cosmology as our fiducial model with $\Omega_{\rm b}=0.0425$,
$\Omega_{\rm c}=0.221$, $\Omega_{\Lambda}=0.737$, $n_{\rm
s}=0.961$, $H_{0}=72.3\hubble$, and $\sigma_{8}=0.834$. This set
of parameter was derived using a joint dataset of {\it WMAP}9 +
SPT + ACT + BAO + $H_{0}$ \cite{Hinshaw12}.

\section{Kinetic SZ power spectrum modeling}
\label{sec:kSZ_model}

\subsection{The kSZ effect}
\label{sec:kSZ_effect}

While traveling from the last scattering surface to us, a
fraction of CMB photons are rescattered by free electrons with a
coherent motion of peculiar velocity along the line-of-sight. The
temperature fluctuations generated by such rescattering is
\cite{Shaw12,Ma02,Zhang04}
\begin{equation}
\frac{\Delta T}{T_{0}}(\hat{n})=\sigma_{\rm T} \int_{0}^{z_{\rm
rei}} \frac{\der z}{(1+z)H(z)}n_{\rm e,i}(z) e^{-\tau(z)}
(\vec{v}\cdot \hat{n}), \label{eq:deltat}
\end{equation}
where $T_{0}\simeq 2.725\text{ } $K is the average temperature of
CMB, $\sigma_{\rm T}$ is the Thomson cross-section for an
electron, $H(z)$, $\tau(z)$ and $n_{\rm e,i}(z)$ are the Hubble
parameter, optical depth and the ionized free-electron number density
respectively, and
$\vec{v}\cdot\hat{n}$ is the peculiar velocity of electrons along
the line-of-sight. We choose the upper limit of the integral to be
$z_{\rm rei}=10$ since we mainly focus on the kinetic SZ effect after the reionization, 
which happens at $z=10$ in our fiducial cosmological model used in this analysis.
Later we will see that the exact kSZ signal is not very
sensitive to this upper limit as long as $z\gtrsim10$.

The optical depth at redshift $z$ is \cite{Shaw12,Ma02,Zhang04}
\begin{eqnarray}
\tau(z)=\sigma_{\rm T} \int^{z}_{0} \frac{c \overline{n}_{\rm
e,i}(z')}{(1+z')H(z')}\der z', \label{eq:optical}
\end{eqnarray}
where $\overline{n}_{\rm e,i}(z)$ is the mean ionized
free-electron number density. If we assume that at $z<z_{\rm rei}$
the hydrogen is completely ionized, then
\cite{Shaw12,Ma02,Zhang04}
\begin{eqnarray}
\overline{n}_{\rm e,i}=\frac{\chi \rho_{\rm g}(z)}{\mu_{\rm
e}m_{\rm p}}, \label{eq:free-ne}
\end{eqnarray}
where $\rho_{\rm g}(z)=\rho_{\rm g,0}(1+z)^{3}$ is the mean gas
density at redshift $z$, $\mu_{\rm e} = 1.14$ is the mean mass per
electron, and
\begin{eqnarray}
\chi=\frac{1-Y_{\rm p}(1-N_{\rm H_{e}}/4)}{1-Y_{\rm p}/2},
\end{eqnarray}
is the fraction of 
ionized electrons. $Y_{\rm
p}=0.24$ is the primordial helium abundance, and $N_{\rm H_{e}}$
is the number of helium electron ionized. We leave the derivation
of Eq.~(\ref{eq:free-ne}) in Appendix~\ref{sec:chi-mu-derive}.

Since the free-electron number density is related to its mean
value by $n_{\rm e,i}= \overline{n}_{\rm e,i}(1+\delta)$, and we
define the density averaged peculiar velocity as the ``momentum
field''\footnote{This definition is widely used in many previous
literatures, e.g.~\cite{Cooray02,Rubino-Martin04}}
$\vec{q}=\vec{v}(1+\delta)$ ($\delta=(\rho-
\overline{\rho})/\overline{\rho}$ is the density contrast), then
Eq.~(\ref{eq:deltat}) becomes
\begin{equation}
\frac{\Delta T}{T_{0}}(\hat{n})=\left(\frac{\sigma_{\rm
T}\rho_{\rm g,0}}{\mu_{\rm e}m_{\rm p}}\right) \int^{z_{\rm
rei}}_{0} \frac{(1+z)^{2}}{H(z)}\chi e^{-\tau(z)}
(\vec{q}\cdot\hat{n}) \der z, \label{eq:deltat2}
\end{equation}

Expanding Eq.~(\ref{eq:deltat2}) onto spherical
harmonics and calculating the angular power spectrum $C_{\ell}$ of the expansion coefficients $a_{\ell m}$, one can obtain the kSZ angular power spectrum \cite{Dodelson95,Jaffe98,Ma02,Shaw12} under the Limber approximation \citep{Limber53},
\begin{eqnarray}
C_{\ell} &= &
\frac{8\pi^{2}}{(2\ell+1)^{3}}\left(\frac{\sigma_{\rm T}\rho_{\rm
g,0}}{\mu_{\rm e}m_{\rm p}} \right)^{2} \int^{z_{\rm
zei}}_{0} (1+z)^{4} \nonumber \\
& \times & \chi^{2}\Delta^{2}_{\rm b}(\ell/x,z)
e^{-2\tau(z)}\frac{x(z)}{c H(z)} \der z, \label{eq:ksz-cls}
\end{eqnarray}
where $x(z)=\int^{z}_{0}(c/H(z'))\der z'$ is the comoving distance
out to redshift $z$, $k=\ell/x$, and $\Delta^{2}_{\rm b}(k,z)$ is
the curl component of the momentum power spectrum at redshift $z$.
The expression for $\Delta^{2}_{\rm b}(k,z)$ is
\cite{Dodelson95,Jaffe98,Ma02,Shaw12},
\begin{widetext}
\begin{eqnarray}
\Delta^{2}_{\rm b}(k,z)  =  \frac{k^{3}}{2\pi^{2}}
\int\frac{\der^{3}\vec{k}'}{(2\pi)^{3}} \left[(1-\mu^{2})P_{\delta
\delta}(|\vec{k}-\vec{k}'|,z) P_{\rm vv}(k',z)
-\frac{(1-\mu^{2})k'}{|\vec{k}-\vec{k}'|}P_{\delta \rm
v}(|\vec{k}-\vec{k}'|)P_{\delta \rm v}(k') \right],
\label{eq:deltab2}
\end{eqnarray}
\end{widetext}
where $P_{\delta \delta}$ ($P_{\rm vv}$) is the linear density
(velocity) power spectrum and $P_{\delta \rm v}$ is the
density-velocity cross spectrum. $\mu=\hat{k}\cdot \hat{k}'$ is
the cosine angle between vectors $\vec{k}$ and $\vec{k}'$. In the
linear theory regime, the continuity equation indicates that the
Fourier space velocity field ($\tilde{v}(\vec{k})$) is related to
density field through \cite{Dodelson03,Sarkar07},
\begin{eqnarray}
\tilde{v}(\vec{k})=if \dot{a}\delta(\vec{k})\frac{\vec{k}}{k^{2}},
\label{eq:fourier-v}
\end{eqnarray}
where $f=\der \log D/\der \log a$, and $D$ is the linear growth
factor. Therefore the peculiar velocity power spectrum and
density-velocity cross-spectrum are related to the linear density
power spectrum as \cite{Shaw12,Dodelson03},
\begin{eqnarray}
P_{\rm vv}(k)=\left(\frac{f\dot{a}}{k} \right)^{2}P_{\delta
\delta}(k); \text{    } P_{\delta \rm
v}(k)=\left(\frac{f\dot{a}}{k} \right)P_{\delta \delta}(k).
\label{eq:Pvv}
\end{eqnarray}
Therefore Eq.~(\ref{eq:deltab2}) becomes \cite{Shaw12,Dodelson03}\footnote{Note that Eq. (\ref{eq:deltab2-OV}) only holds in the models where the growth is scale-independent. For more general cases in which the growth is scale-dependent, {\it e.g.} the models with massive neutrinos or the modified gravity models, one should leave the function $f$, which is a function of $k$ and $z$, inside the integral.},
\begin{eqnarray}
\Delta^{2}_{\rm b}(k,z) & = & \frac{k^{3}}{2\pi^{2}}(\dot{a}f)^{2}
\int\frac{\der^{3}\vec{k}'}{(2\pi)^{3}} P_{\delta
\delta}(|\vec{k}-\vec{k}'|) \nonumber \\ & \times & P_{\delta
\delta}(k') I(k,k') , \label{eq:deltab2-OV}
\end{eqnarray}
where
\begin{eqnarray}
I(k,k')=\frac{k(k-2k' \mu)(1-\mu^{2})}{k'^{2}(k^{2}+k'^{2}-2 k
k'\mu)}, \label{eq:kernel}
\end{eqnarray}
is the kernel function that couples linear velocity field with
density field.

Therefore by substituting Eq.~(\ref{eq:kernel}) into
Eq.~(\ref{eq:deltab2-OV}) and combining with
Eq.~(\ref{eq:ksz-cls}), one can obtain the power spectrum of
kinetic SZ effect, aka Ostriker-Vishniac effect (hereafter OV
effect) \cite{Ostriker86}, which corresponds to the case where the
CMB photons are rescattered by linear structure of galaxy
clusters through the linear velocity modes (such as the bulk motion).

On the other hand, the nonlinearity of the structure formation can affect the kSZ power spectrum significantly on scales of $\ell>1000$.
Refs. \cite{Ma02,Hu00,Zhang04} demonstrate that the full kSZ effect is determined by
the non-linear matter density field $P^{\rm NL}_{\delta \delta}$
cross-correlating with the linear velocity field. 
One can correct for the nonlinearity by
replacing the linear matter power spectrum $P_{\delta \delta}$ in
Eq.~(\ref{eq:deltab2-OV}) with non-linear matter power spectrum
$P^{\rm NL}_{\delta \delta}$, \cite{Shaw12}, {\it i.e.},
\begin{eqnarray}
\Delta^{2}_{\rm b}(k,z) & = & \frac{k^{3}}{2\pi^{2}}(\dot{a}f)^{2}
\int\frac{\der^{3}\vec{k}'}{(2\pi)^{3}} P^{\rm NL}_{\delta
\delta}(|\vec{k}-\vec{k}'|) \nonumber \\
& \times & P_{\delta \delta}(k') I(k,k') . \label{eq:deltab2-NL}
\end{eqnarray}
In addition, there is no need to replace linear velocity field
with non-linear velocity field. This is because velocity power
spectrum has an extra $1/k^{2}$ factor than the matter power
spectrum, so there is more weight on larger scales than the matter
power spectrum. Therefore it turns out that this extra factor make
the velocity field rather insensitive to the small scale
non-linear behavior \cite{Ma02}.

Throughout this paper, we calculate the linear and non-linear
matter power spectrum using the public code {\tt CAMB}
\cite{CMABweb} which automatically incorporates the {\tt HALOFIT}
\cite{Smith03,Takahashi12} prescription for the
non-linear matter power spectrum.

\subsection{Gaseous pressure}
\label{sec:gas}

\begin{figure*}[tbp]
\centerline{
\includegraphics[bb=0 0 631 391, width=3.5in]{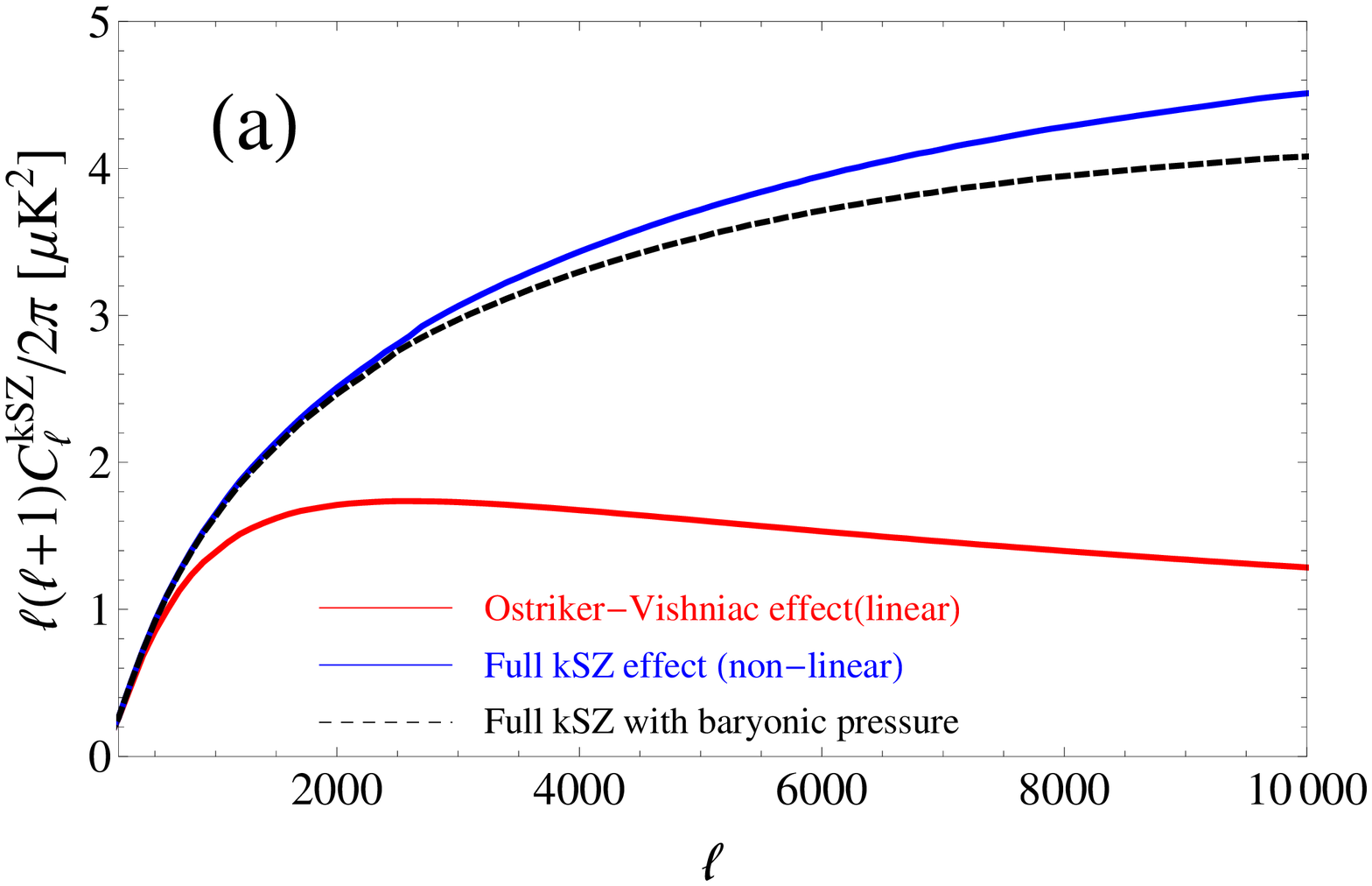}
\includegraphics[bb=0 0 737 416, width=3.9in]{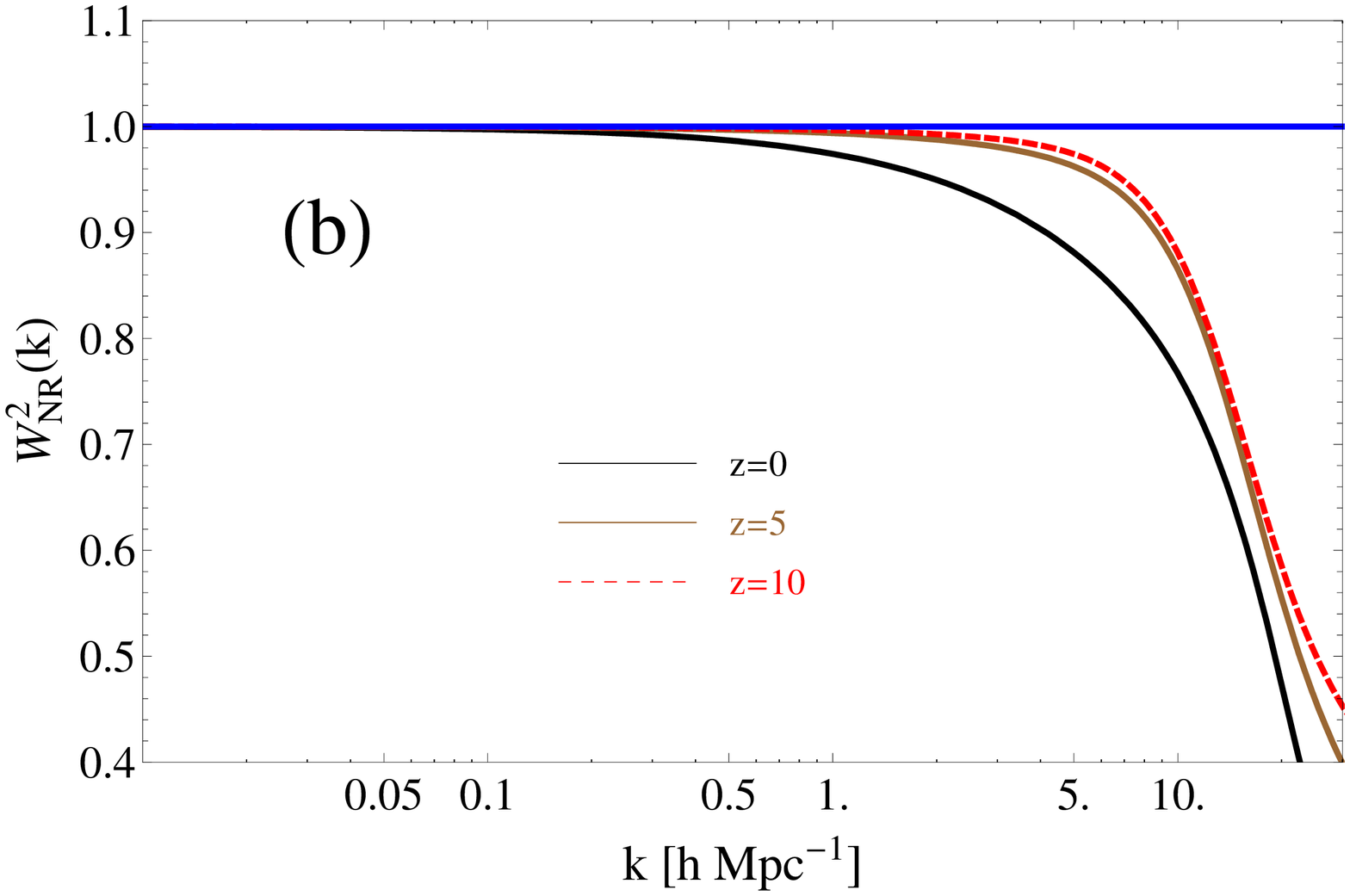}}
\caption{\textit{Panel(a)}: Kinetic SZ effect angular power
spectrum $D_{\ell}=\ell(\ell+1)C_{\ell}/2\pi$. Red solid, blue
solid and black dashed line corresponds to the Ostriker-Vishniac
(OV) effect, full non-linear kSZ effect and the full effect with
gas pressure; \textit{Panel (b)}: Window function of gas pressure
at different redshifts (Eq.~(\ref{eq:wl-simu})).}
\label{fig:ksz-standard}
\end{figure*}

In the kSZ power spectrum calculations, it is
commonly assumed that the density distribution of the baryonic gas
follows exactly that of dark matter, so there is no ``bias'' in
between $\delta_{\rm gas}$ and $\delta_{\rm DM}$
\cite{Dodelson95,Jaffe98,Ma02,Zhang04}. However, on small scales,
a significant fraction of baryons are in form of gas, thus
the thermal pressure of baryons can erase the density fluctuations
in the gas distribution on small scales \cite{Shaw12}. This
``suppression'' effect can be modeled as a window function $W(k)$
such that \cite{Shaw12},
\begin{eqnarray}
P^{\rm NL}_{\rm gas}(k,z)=W^{2}(k,z)P^{\rm NL}_{\rm DM}(k,z).
\label{eq:wl}
\end{eqnarray}
Here we use the fitting formula of $W(k)$ developed by
\cite{Shaw12},
\begin{eqnarray}
W^{2}(k,a)=\frac{1}{2}\left\{e^{-k/k_{\rm f}}+\frac{1}{1+
[1+g(a)k/k_{\rm f}]^{7/2}} \right\}, \label{eq:wl-simu}
\end{eqnarray}
where the filter scale $k_{\rm f}=12.6/a+6.3$ and $g(a)=0.84/a$.
This fitting formula is proved to provide a better fit to the gas
density power spectrum than the analytic formula developed by
\citet{Gnedin98} through the comparison with ``BolshoiNR and
L60N'' numerical simulations shown in \cite{Shaw12}. Thus, by
incorporating the gas pressure window function, the power spectrum
$\Delta^{2}_{\rm b}(k,z)$ becomes \cite{Shaw12},
\begin{eqnarray}
\Delta^{2}_{\rm b}(k,z) & = & \frac{k^{3}}{2\pi^{2}}(\dot{a}f)^{2}
\int\frac{\der^{3}\vec{k}'}{(2\pi)^{3}} W^{2}(|\vec{k}-\vec{k}'|,z) \nonumber \\
& \times &  P^{\rm NL}_{\delta \delta}(|\vec{k}-\vec{k}'|)
P_{\delta \delta}(k') I(k,k') . \label{eq:deltab2-NL2}
\end{eqnarray}
Note that we assume that the velocity of gas
follows exactly the velocity of dark matter, so there is no
velocity bias between them.

In Fig.~\ref{fig:ksz-standard}, we plot the kSZ angular power
spectrum and gas window function in panels~(a) and (b) respectively. In
Fig.~\ref{fig:ksz-standard} (b), we can see that on the large scales,
$W(k)\simeq 1$ while on small scales $W(k) \rightarrow 0$ as $k$
increases due to the gas thermal pressure force. The suppression
is not very significant at the onset of the gravitational collapse
(high $z$), but as structures gradually collapse, the suppression
propagates progressively to larger and larger scales. In
Fig.~\ref{fig:ksz-standard} (a), we plot the kSZ angular power spectrum $D_{\ell}\equiv \ell (\ell+1)C_{\ell}/2\pi$ as a
function of $\ell$. One can see that the linear OV effect only
produces the signal peaking at $\ell \simeq 2000$, and gradually
decreases at higher $\ell$. This is because the linear
perturbation is sensitive to linear modes which are
generically on large scales. One the other hand, using non-linear
matter power spectrum instead (Eq.~(\ref{eq:deltab2-NL})) to calculate the
full kSZ effect, 
one obtains the blue solid line, whose amplitude is about $3$
times higher than that of the linear one on small scales. The
$D_{\ell}$ of the full kSZ power spectrum is about $3.06 \mu
\textrm{K}^{2}$ on scales of $\ell \sim 3000$. In addition, if we
incorporate the
window function 
to account for the fact that a fraction of density fluctuations will be suppressed by the gaseous pressure on small scales, {\it i.e.} to use Eq.~(\ref{eq:deltab2-NL2}), the total signal drops by a
factor of $3\%$--$10\%$ on $\ell \sim 2000$ to $10000$.

In order to see clearly how dark energy affects the kSZ signals,
in the following analysis, we will adopt the full non-linear kSZ
effect without gas pressure as our default model, and discuss the
effect of dark energy on this full-kSZ signal. Of-course, when
using this model to compare with observations, one needs to
consider the effect of gas pressure, which can only be well
understood from numerical simulations.


\section{Dark energy imprints}
\label{sec:dark_energy}

\begin{figure*}[tbp]
\centerline{
\includegraphics[bb=0 0 650 425, width=3.2in]{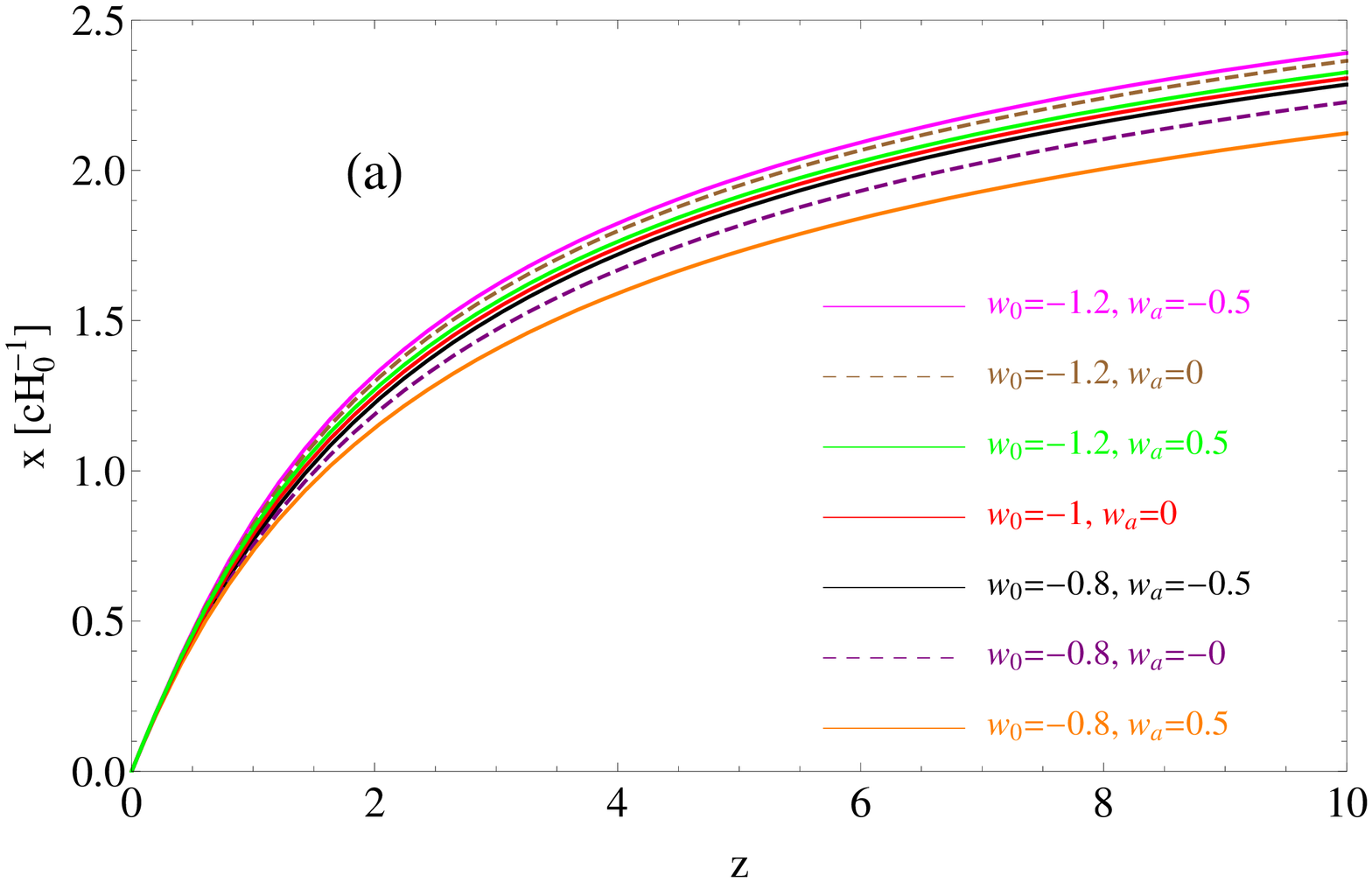}
\includegraphics[bb=0 0 630 404, width=3.2in]{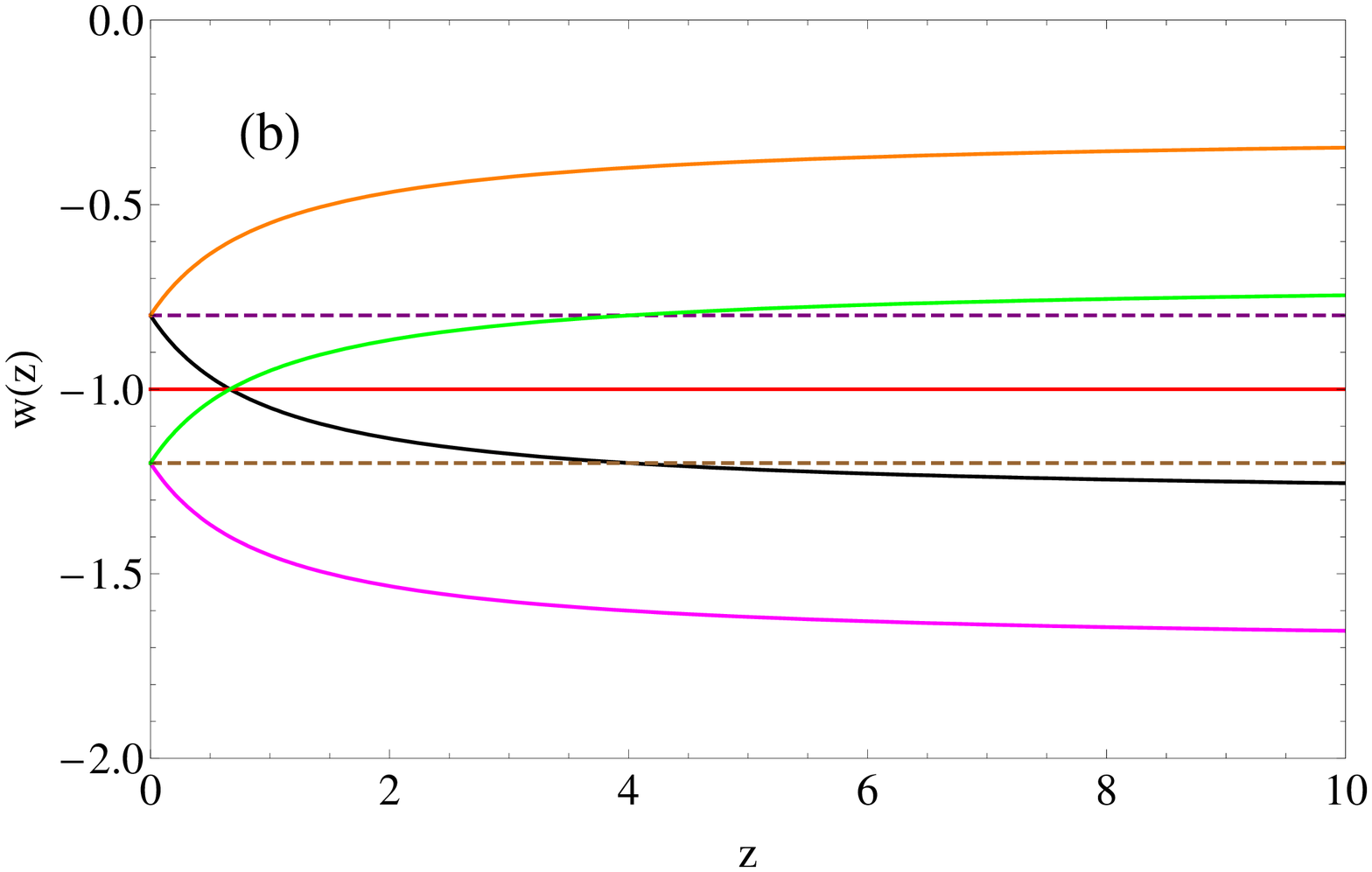}}
\caption{\textit{Panel(a)}: Comoving distance
$x=\int^{\infty}_{0}(c/H(z'))\der z'$ for seven dark energy models
with different parameters of EoS. \textit{Panel (b)}: the
evolution of EoS for seven dark energy models. The colour scheme
of the two panels is shown on panel~(a).} \label{fig:x-wz}
\end{figure*}

\begin{figure*}[tbp]
\centerline{
\includegraphics[bb=0 0 471 316, width=3.2in]{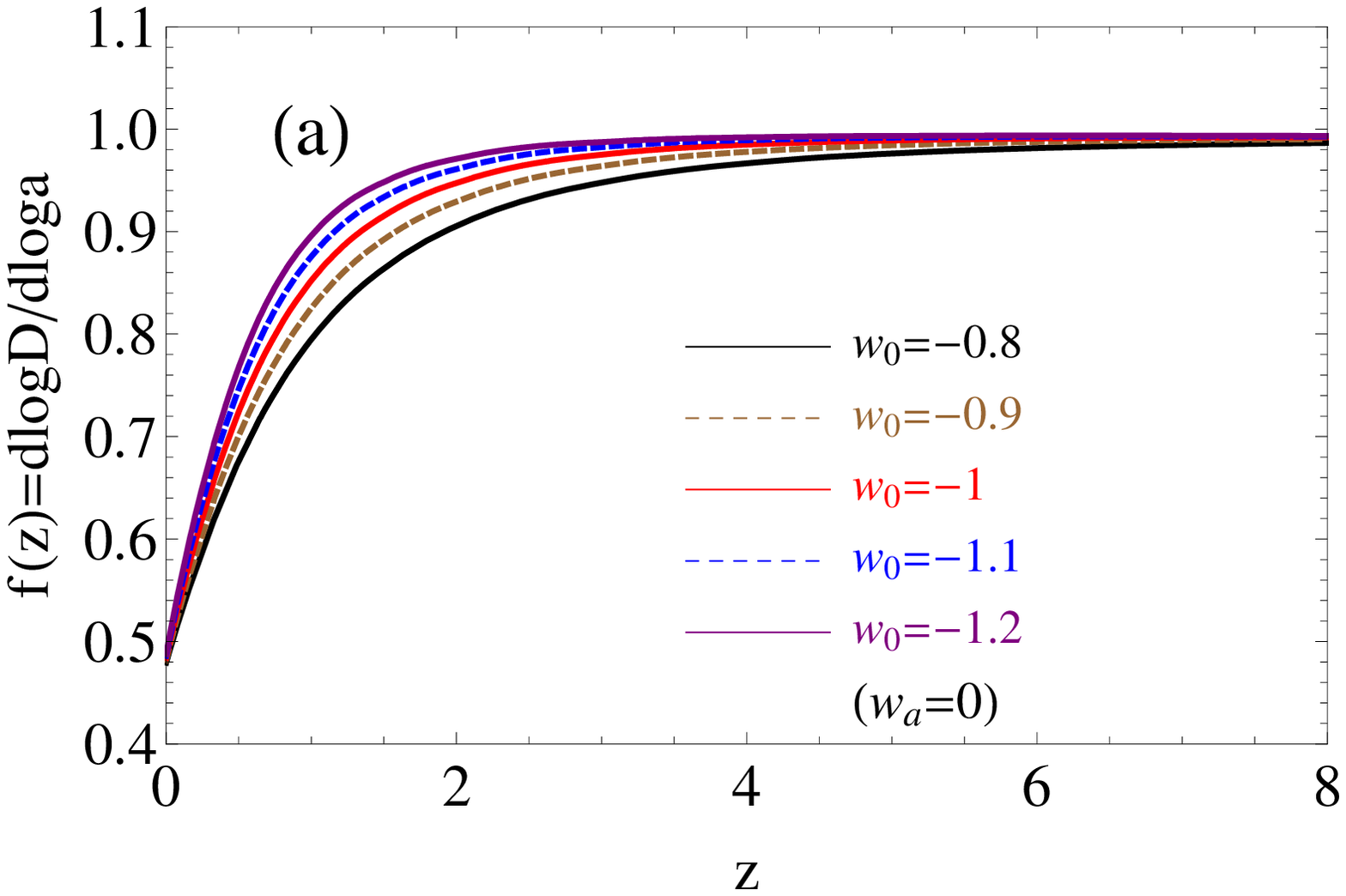}
\includegraphics[bb=0 0 617 402, width=3.2in]{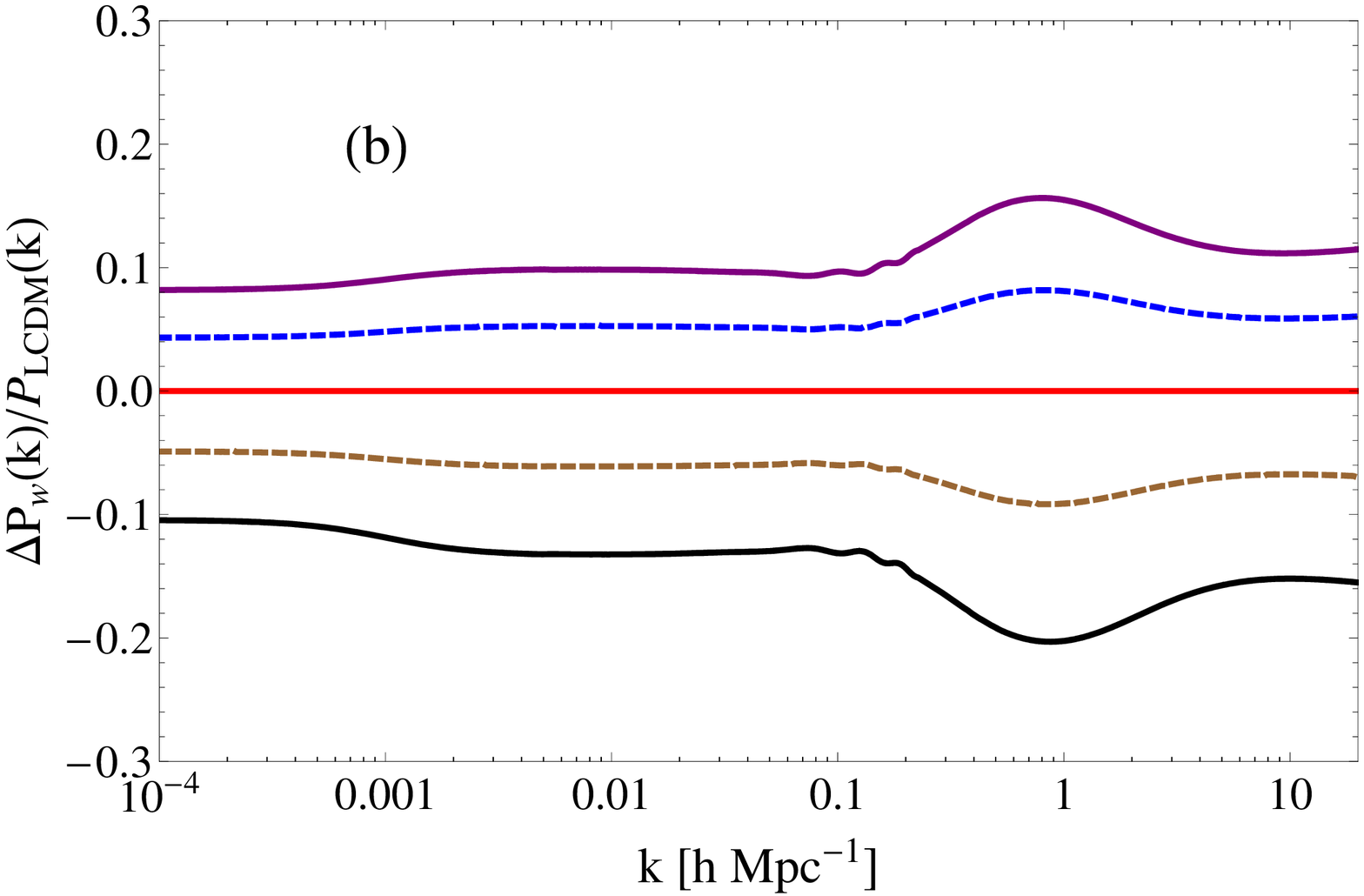}}
\caption{\textit{Panel(a)}: Growth function $f(z)=\der
\log(D)/\der \log (a)$ for the five dark energy models with
constant equation of state parameters. \textit{Panel (b)}: the
fractional difference of power spectrum $\Delta P_{w}(k)/P_{\rm
LCDM}(k)=(P_{w}(k)-P_{\rm LCDM}(k))/P_{\rm LCDM}(k)$ between dark
energy models with four different $w$ values and $\Lambda$CDM
model. The colour scheme of the two panels is shown on panel~(a).}
\label{fig:deltaP1-fz1}
\end{figure*}

\begin{figure*}[tbp]
\centerline{
\includegraphics[bb=0 0 574 379, width=3.2in]{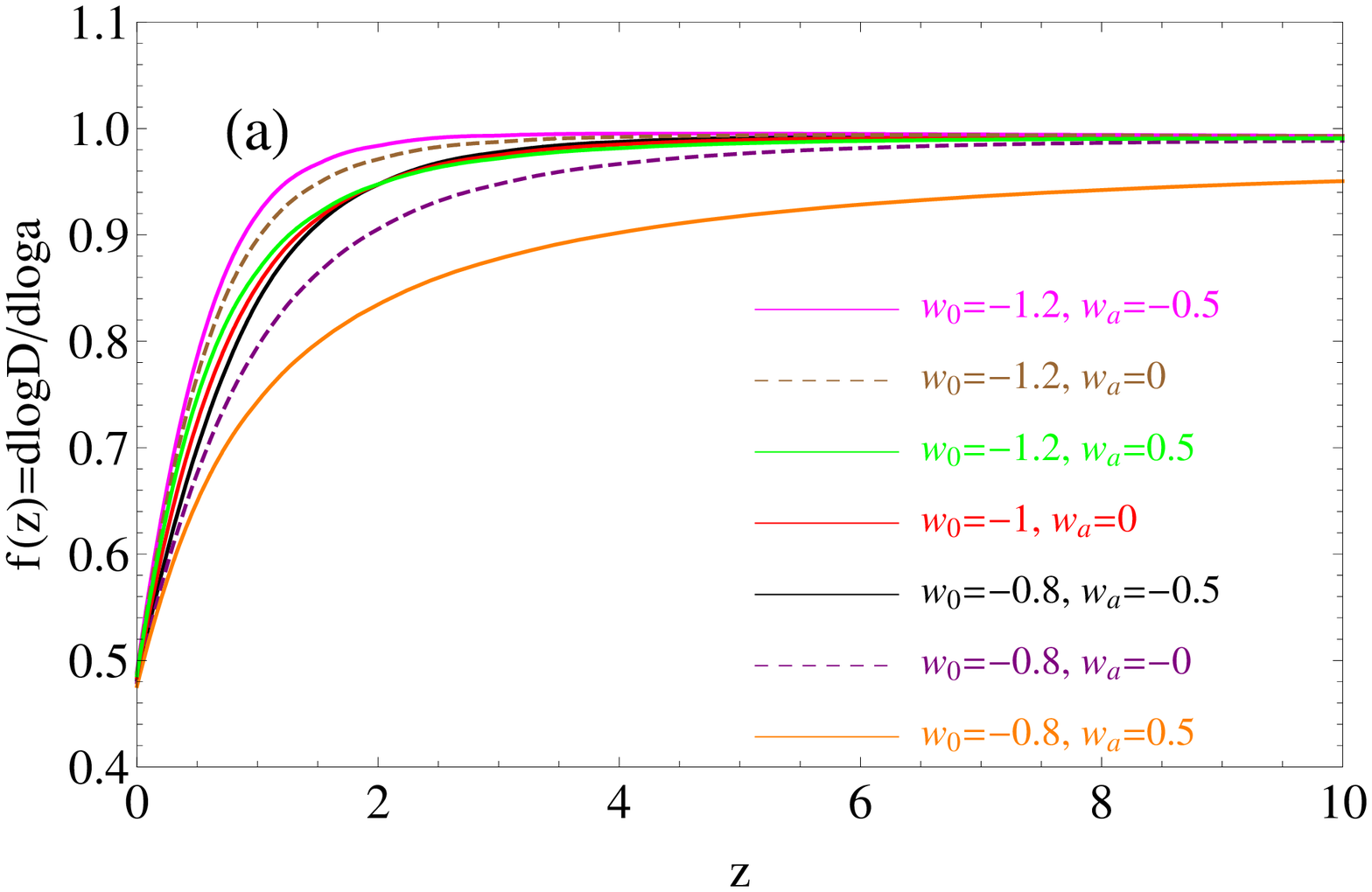}
\includegraphics[bb=0 0 614 398, width=3.2in]{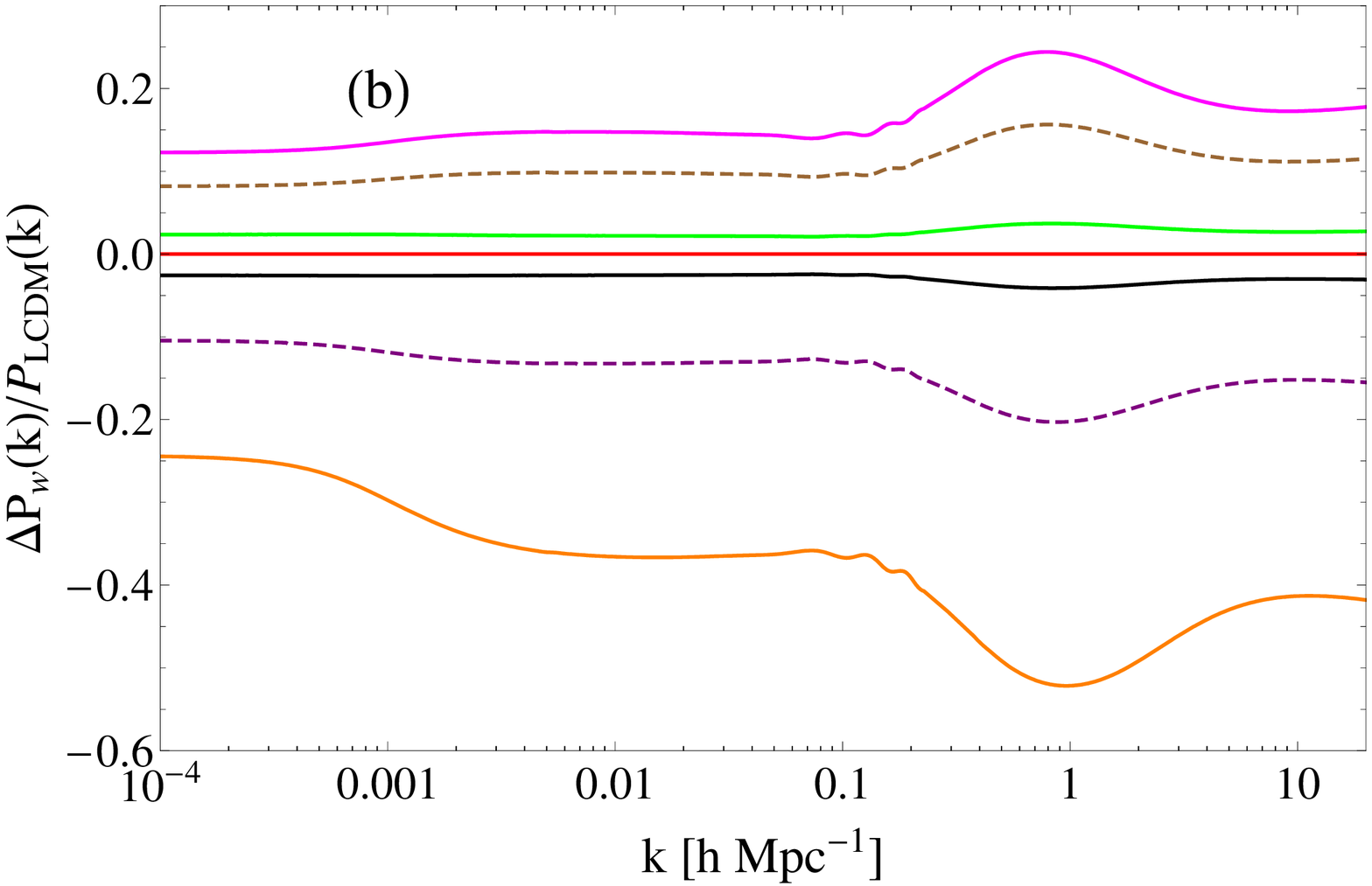}}
\caption{\textit{Panel(a)}: Growth function $f(z)=\der
\log(D)/\der \log (a)$ for the seven dark energy models with
time-varying EoS parameters ($w(z)=w_{0}+w_{a}z/(1+z)$).
\textit{Panel (b)}: the fractional difference of power spectrum
$\Delta P_{w}(k)/P_{\rm LCDM}(k)=(P_{w}(k)-P_{\rm LCDM}(k))/P_{\rm
LCDM}(k)$ between dark energy models with six different $w(z)$
evolution and $\Lambda$CDM model. The colour scheme of the two
panels is shown on panel~(a).} \label{fig:deltaP2-fz2}
\end{figure*}

In this section we shall first review how the dark energy EoS
changes the comoving distance $x(z)$, and then show how the
time-varying dark energy affects the structure growth, and
eventually we analyze how the kSZ power spectrum is affected by
dark energy.

\subsection{ EoS $w(z)$ and comoving distance $x(z)$}
\label{sec:wz}

We adopt the Chevallier-Polarski-Linder (CPL) parametrization
\cite{Chevallier01,Linder03} of dark energy, {\it i.e.,}
$w(a)=w_{0}+w_{a}(1-a)$ where $w_{0}$ and $w_{a}$ are the two free
parameters \cite{Hinshaw12}. In this parametrization form, the
fractional matter density and dark energy density evolve as
\begin{eqnarray}
\Omega_{\rm m}(z) &=& \Omega_{\rm m}^0 (1+z)^{3}, \nonumber \\
\Omega_{\rm DE}(z) &=& \Omega_{\rm DE
}^0(1+z)^{3(1+w_{0}+w_{a})}\exp\left(-\frac{3w_{a}z}{1+z}
\right),\label{eq:omgm-la-CPL}
\end{eqnarray}
where $\Omega_{\rm m}^0=\Omega_{\rm c}^0+\Omega_{\rm b}^0$ and
$\Omega_{\rm DE}^0$ are the matter and dark energy density at
present time and their values are set to be the default values in
Sec.~\ref{sec:intro}. The we can substitute these two equations
into the Friedmann equation $H(z)=H_{0}[\Omega_{\rm
m}(z)+\Omega_{\rm DE}(z)]^{1/2}$ to calculate the Hubble
expansion and comoving distance $x(z)$.

In the following analyses, we take representative values of
$w_{0}$ to be $-0.8$, $-0.9$, $-1$, $-1.1$ and $-1.2$, and $w_{a}$
of $-0.5$, $0$ and $0.5$. 
All these models are allowed by the joint constraints using
\textit{WMAP}9+SPT+ACT+BAO+$H_{0}$ \cite{Lambdaweb}.

In Fig.~\ref{fig:x-wz}, we plot the dark energy EoS in panel~(b)
and the corresponding comoving distance at redshift $z$ in
panel~(a). One can see that the comoving distance increases as
$w_0$ or $w_a$ drops and {\it vice versa}.
This is simply because a more negative $w_0$ or $w_a$ means a smaller Hubble parameter in the past, thus a larger comoving distance.
This is apparent in Fig.~\ref{fig:x-wz} (a).


This brings up the question of degeneracy. If $w_{0}$
is more negative but $w_{a}$ is positive, this will produce the
similar effect with a less negative $w_{0}$ but more negative
$w_{a}$. For instance, in Fig.~\ref{fig:x-wz}a, we can see that
the $x(z)$ function for $w_{0}=-0.8, w_{a}=-0.5$ is very close to
the model $w_{0}=-1.2, w_{a}=0.5$, and also close to the
$\Lambda$CDM model ($w_{0}=-1$, $w_{a}=0$). This is because the
comoving distance is an integrated effect, although the evolution
of $w(z)$ are different for these models, their integrated effects
are close to each other. This degenerates between the
time-evolving EoS parameters is what we should be aware of when
analyzing the kSZ effect signals.

\subsection{Growth function $f(z)$}
\label{sec:growth}

In the 3D power spectrum of kSZ effect
(Eq.~(\ref{eq:deltab2-NL2})), $\Delta_{\rm b}(k,z)$ function
depends on the evolution of structure growth function $f(z)$, and
also the (non)linear matter power spectrum. The growth function
$f(z)$ is the logarithmic derivative of the growth rate $D(z)$
($\delta(t)=D(t)\delta_{0}$), {\it i.e.} $f(z)=\der \log(D)/\der
\log(a)$.

We use the numerical code  \textsc{camb} \cite{CMABweb} to
calculate the growth function $f$ for various dark energy models
in question. Since \textsc{camb} does not output growth function
directly, we first modify its subroutine and output the density
contrast as a function of redshift, and the calculate its
logarithmic derivative to obtain the growth function. Note that we
included the dark energy perturbation consistently in the
calculation and pay particular attention to the quintom scenario
\cite{quintom} in which $w$ crosses $-1$ during evolution using
the prescription in Ref. \cite{DEP}.

In Fig.~\ref{fig:deltaP1-fz1} (a), we vary the $w_{0}$ value from
$-0.8$ to $-1.2$ while fixing $w_{a}=0$, while in Fig.~\ref{fig:deltaP2-fz2} (a), we vary $w_a$ as well.
One can see that the
$f(z)$ function for various models converge at both ends, say, at
$z=0$ and $z=10$. 
This is easy to understand since $f(z)\simeq\Omega_{\rm m}(z)^{\gamma}$ where $\gamma$ has a weak dependence on $w(z)$. At low $z$, $\Omega_{\rm m}(z)\simeq\Omega_{\rm m0}$ while at the high $z$ end, $\Omega_{\rm m}(z)\simeq1$.
Therefore different values
of $w_{0}$ or $w_a$ mainly affect the evolution in the middle. A more negative $w_{0}$ or $w_a$ makes dark energy less important in the past, which effectively gives structures more time to grow before diluted, thus a larger growth rate. 



\subsection{Power spectrum $P(k)$}
\label{sec:power}

We now compare the power spectrum $P(k)$ in different dark energy
models.

In Fig.~\ref{fig:deltaP1-fz1} (b) and Fig.~\ref{fig:deltaP2-fz2}
(b), we plot the fractional difference of $P(k)$ for different
dark energy models with respect to the
fiducial $\Lambda$CDM model using the same
as in panel~(a). One can see that a more
negative $w_{0}$ or $w_a$,
results in a higher $P(k)$ due to a higher growth rate as discussed.

One can also see a small bump on scales of $k \sim 1 h {\rm
Mpc}^{-1}$, which is due to nonlinearity. For both $w$CDM and
$\Lambda$CDM models, there is an enhancement on $P(k)$ on
quasi-nonlinear scales (e.g., $k \sim 0.1 h {\rm Mpc}^{-1}$) due
to the transition from the 2-halo to 1-halo terms. Since this
transition scale depends on cosmology, a bump structure can appear
on the fractional difference of $P(k)$ between different
cosmological models. Another example of such bumps on the same
scales can be found in fig.~7 of Ref.~\cite{Zhao11}, in where
$\Delta P/P$ is shown for LCDM cosmology with different values of
$\sigma_8$.

\section{kSZ signal for different dark energy models}
\label{sec:kSZ_DE}

\begin{figure}[tbp]
\centerline{
\includegraphics[bb=0 0 503 333, width=3.2in]{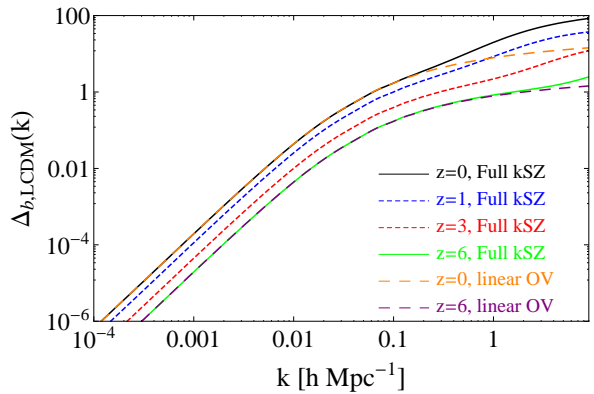}}
\caption{Curl component of momentum power spectrum $\Delta_{\rm
b}(k)$ of $\Lambda$CDM model at four different redshifts. The
linear OV stands for Ostriker-Vishniac effect (replacing $P^{\rm
NL}_{\delta \delta}$ with linear $P_{\delta \delta}$ in
Eq.~(\ref{eq:deltab2-NL2})), while the full kSZ corresponds to
full non-linear results (using non-linear $P^{\rm NL}_{\delta
\delta}$ in Eq.~(\ref{eq:deltab2-NL2})).} \label{fig:deltab-lcdm}
\end{figure}

\begin{figure*}[tbp]
\centerline{
\includegraphics[bb=0 0 605 400, width=3.2in]{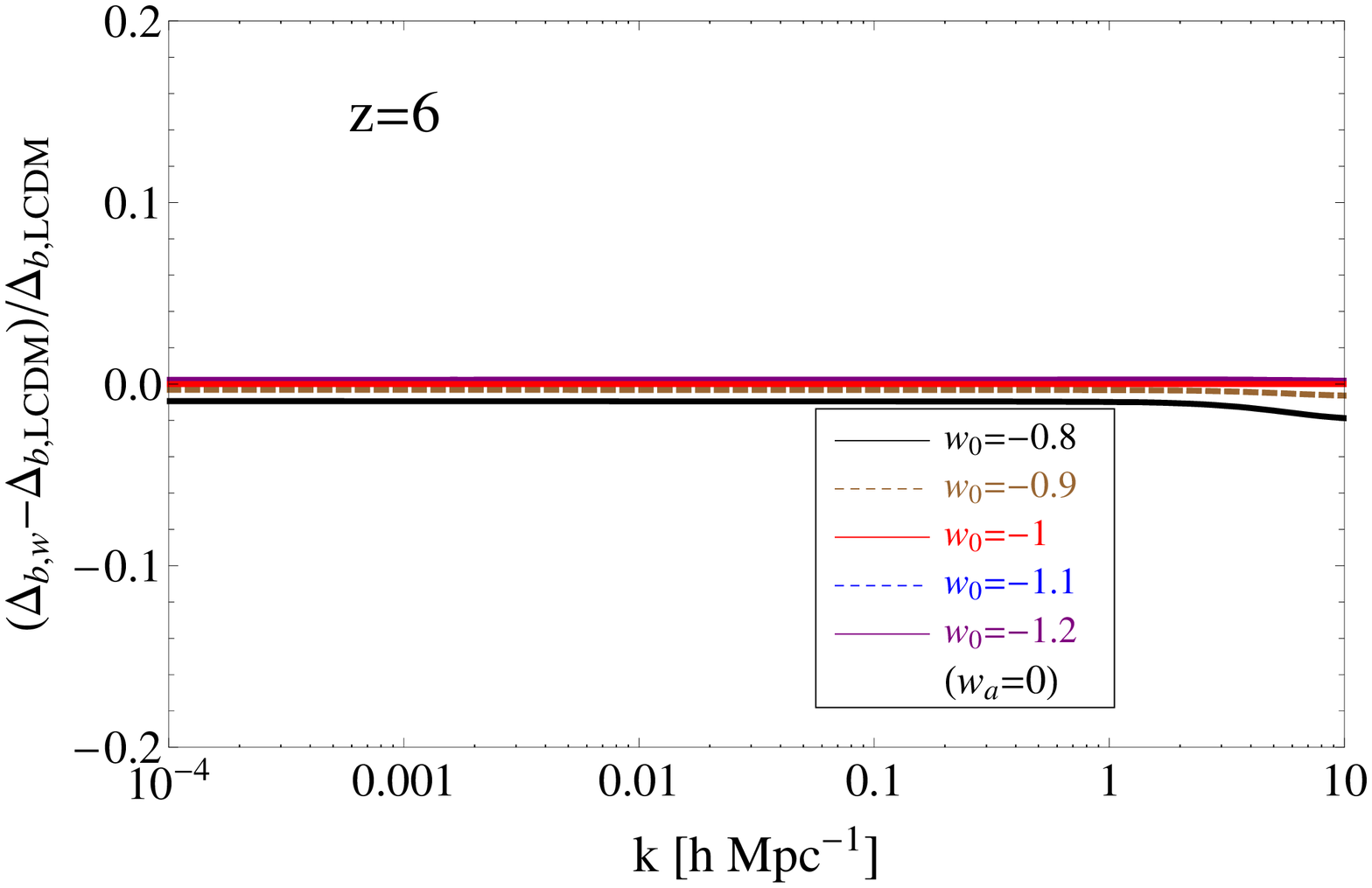}
\includegraphics[bb=0 0 560 364, width=3.2in]{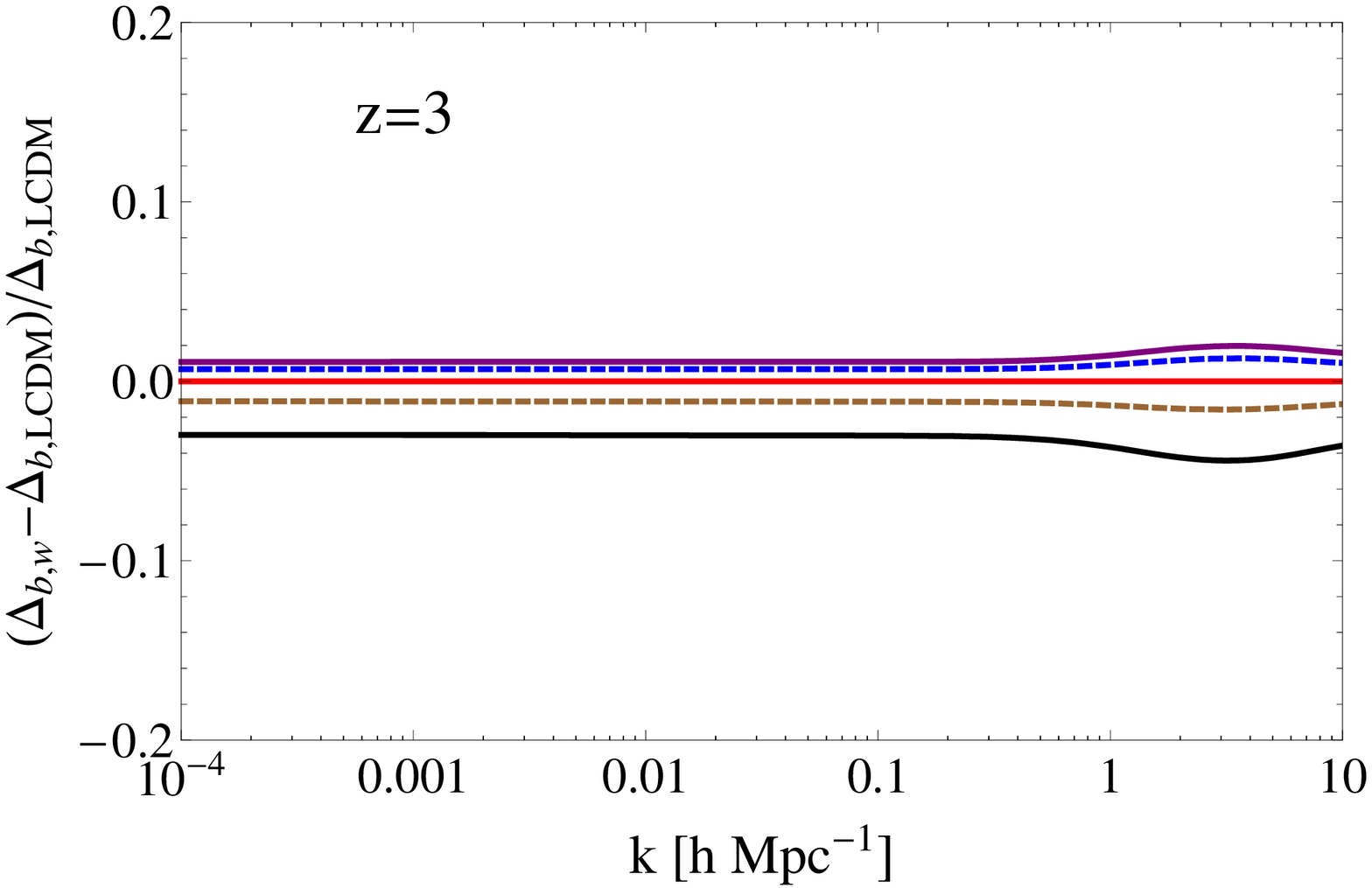}}
\centerline{
\includegraphics[bb=0 0 560 364, width=3.2in]{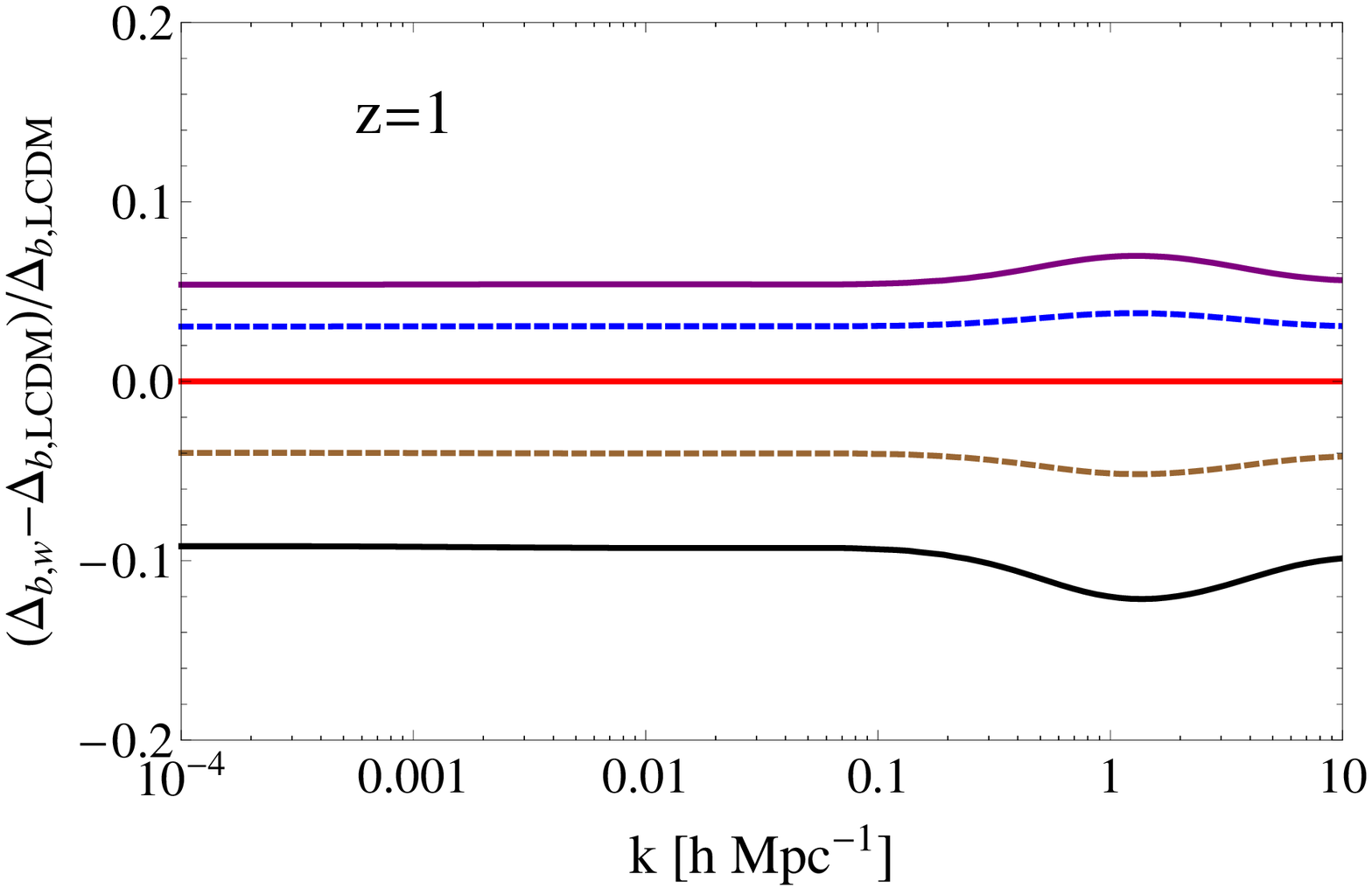}
\includegraphics[bb=0 0 560 364, width=3.2in]{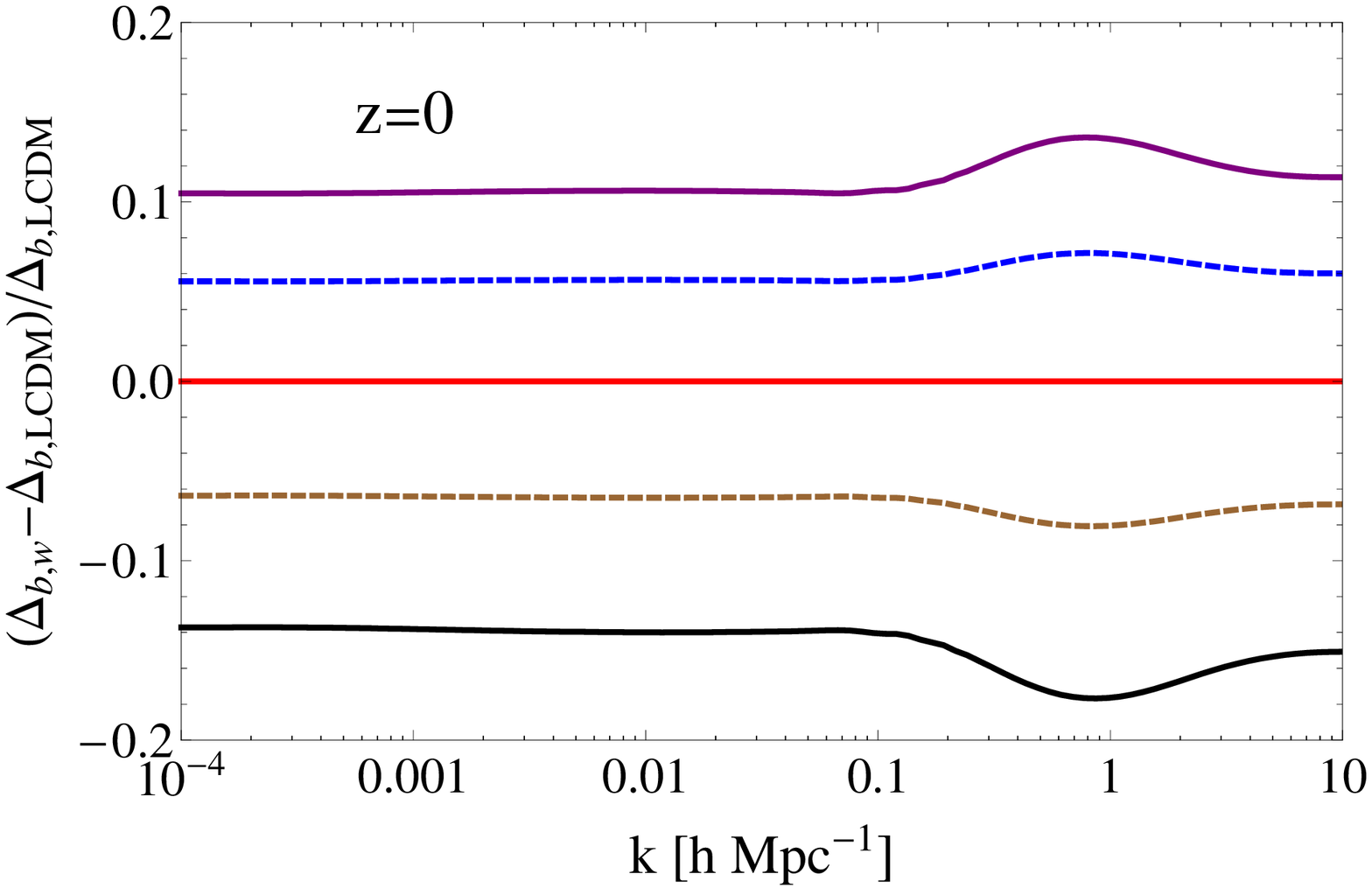}}
\caption{Fractional difference between the power spectrum of curl
momentum field $\Delta_{\rm b}(k)$ at four different redshifts.
The colour scheme for five different dark energy models with
constant EoS is shown on the ``$z=6$'' panel.}
\label{fig:deltab-w0}
\end{figure*}

\begin{figure*}[tbp]
\centerline{
\includegraphics[bb=0 0 735 473, width=3.2in]{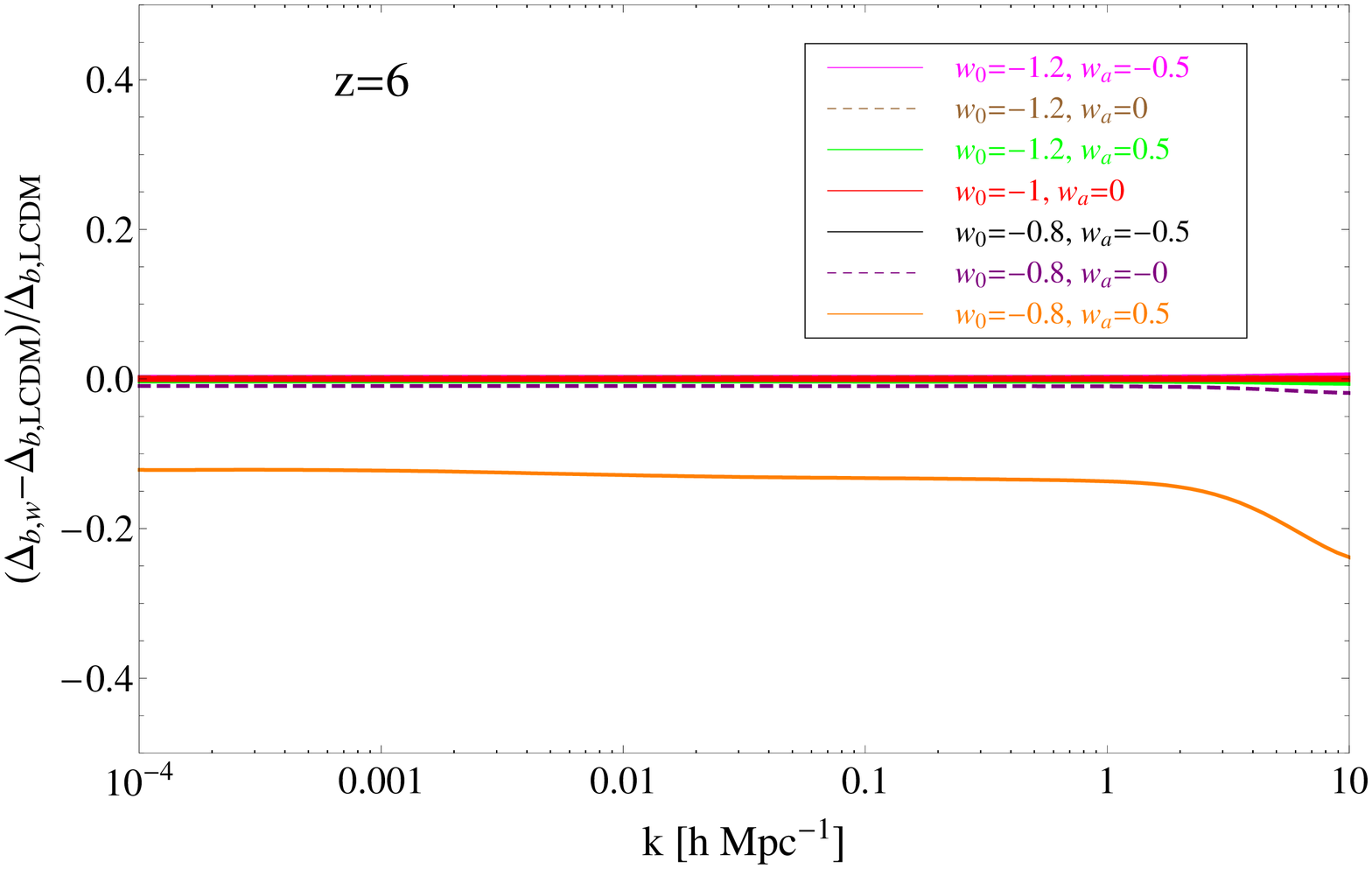}
\includegraphics[bb=0 0 560 364, width=3.2in]{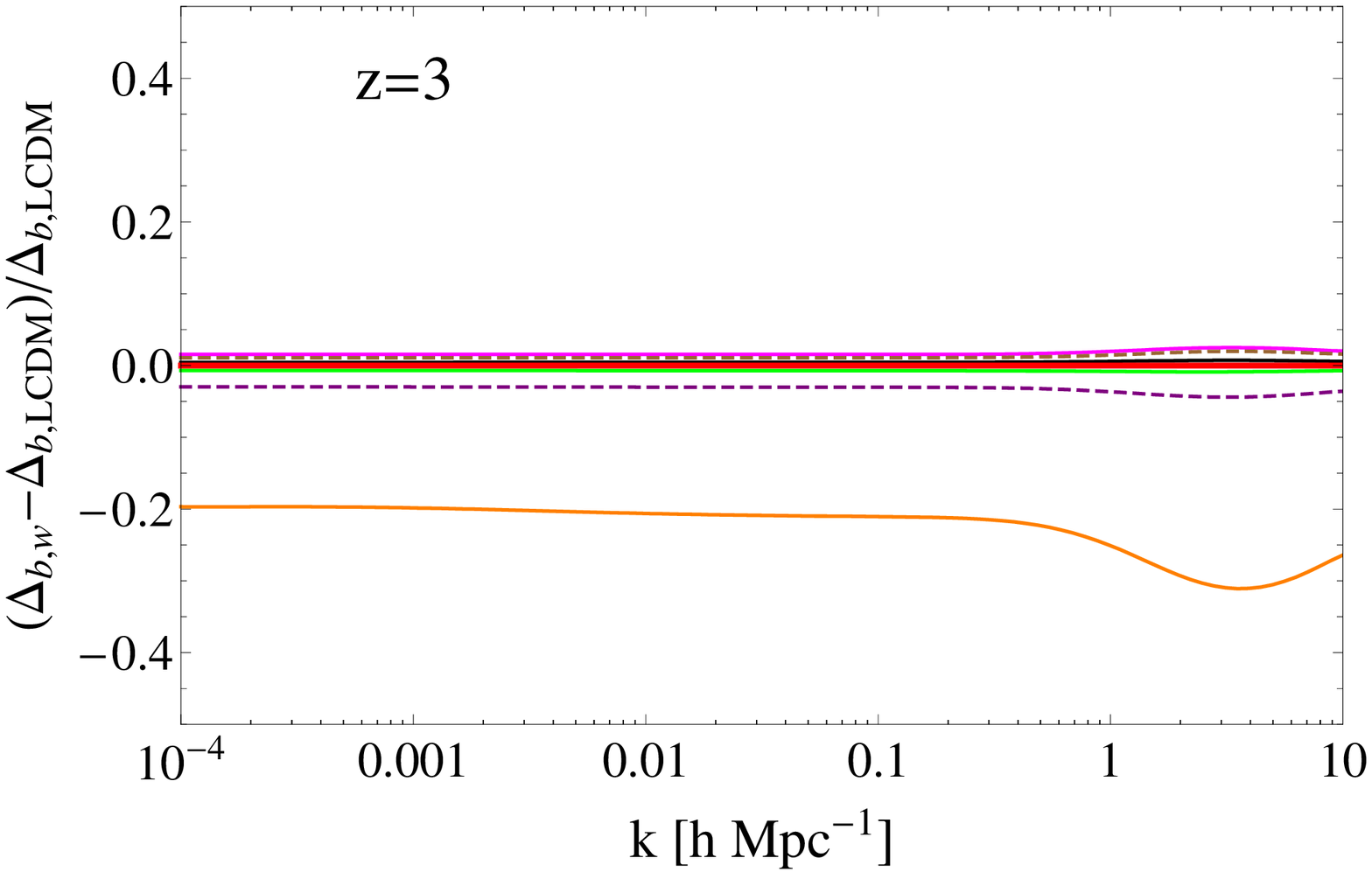}}
\centerline{
\includegraphics[bb=0 0 560 364, width=3.2in]{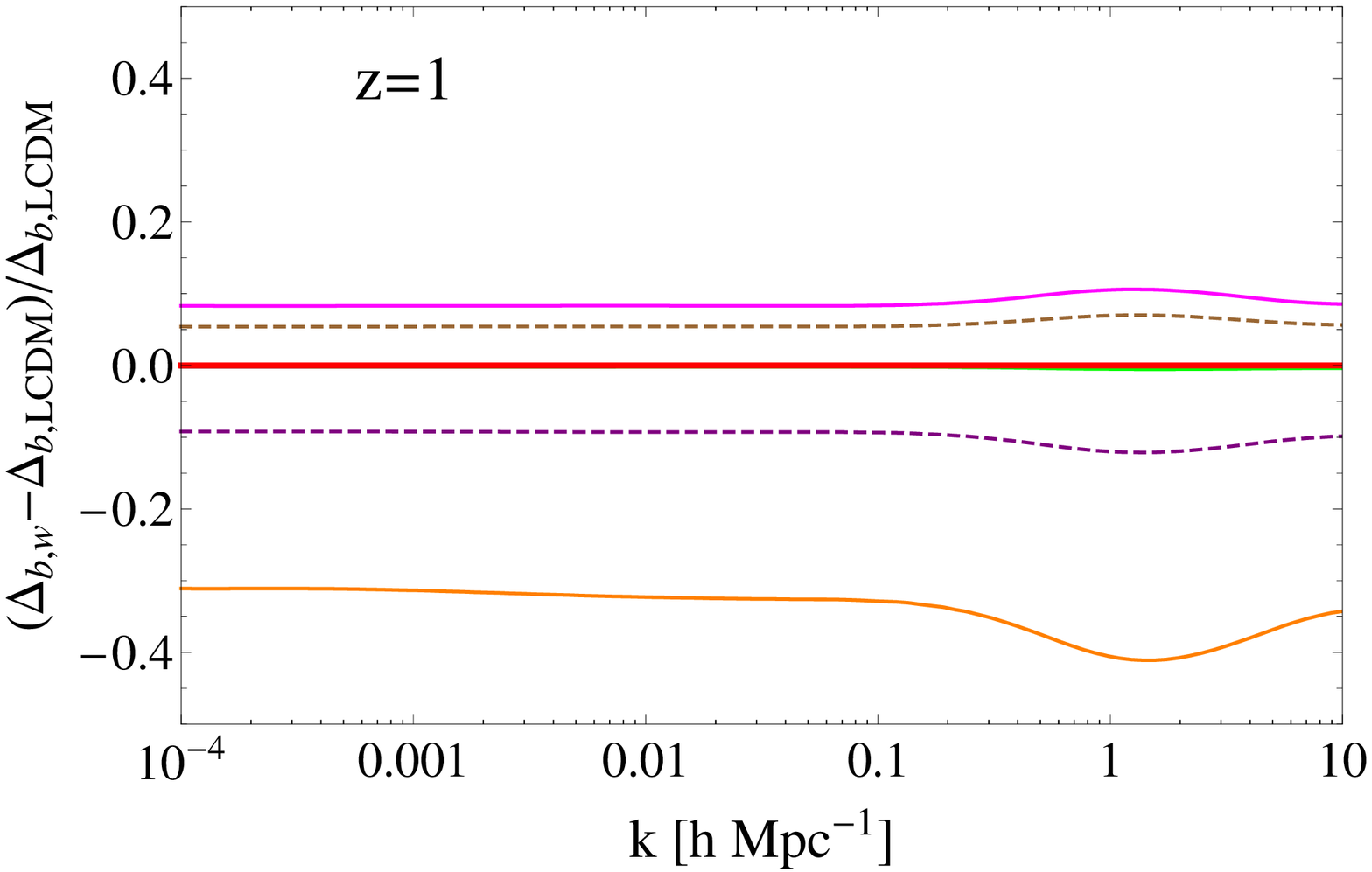}
\includegraphics[bb=0 0 560 364, width=3.2in]{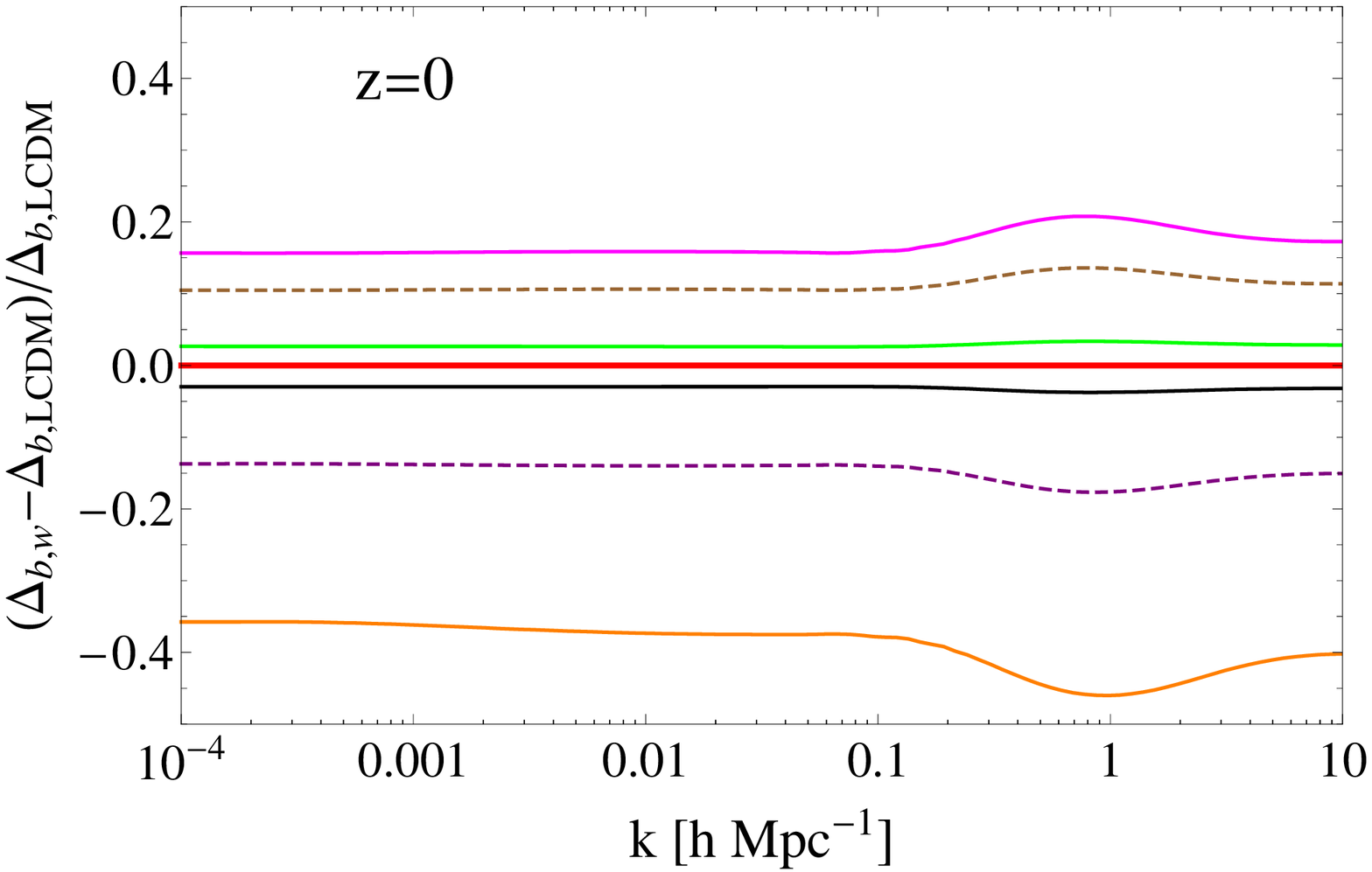}}
\caption{Fractional difference between the power spectrum of
momentum field $\Delta_{\rm b}(k)$ at four different redshifts.
The colour scheme for seven different dark energy models with
time-varying EoS is shown on the ``$z=6$'' panel.}
\label{fig:deltab-w0-wa}
\end{figure*}

\begin{figure*}[tbp]
\centerline{
\includegraphics[bb=0 0 710 470, width=3.1in]{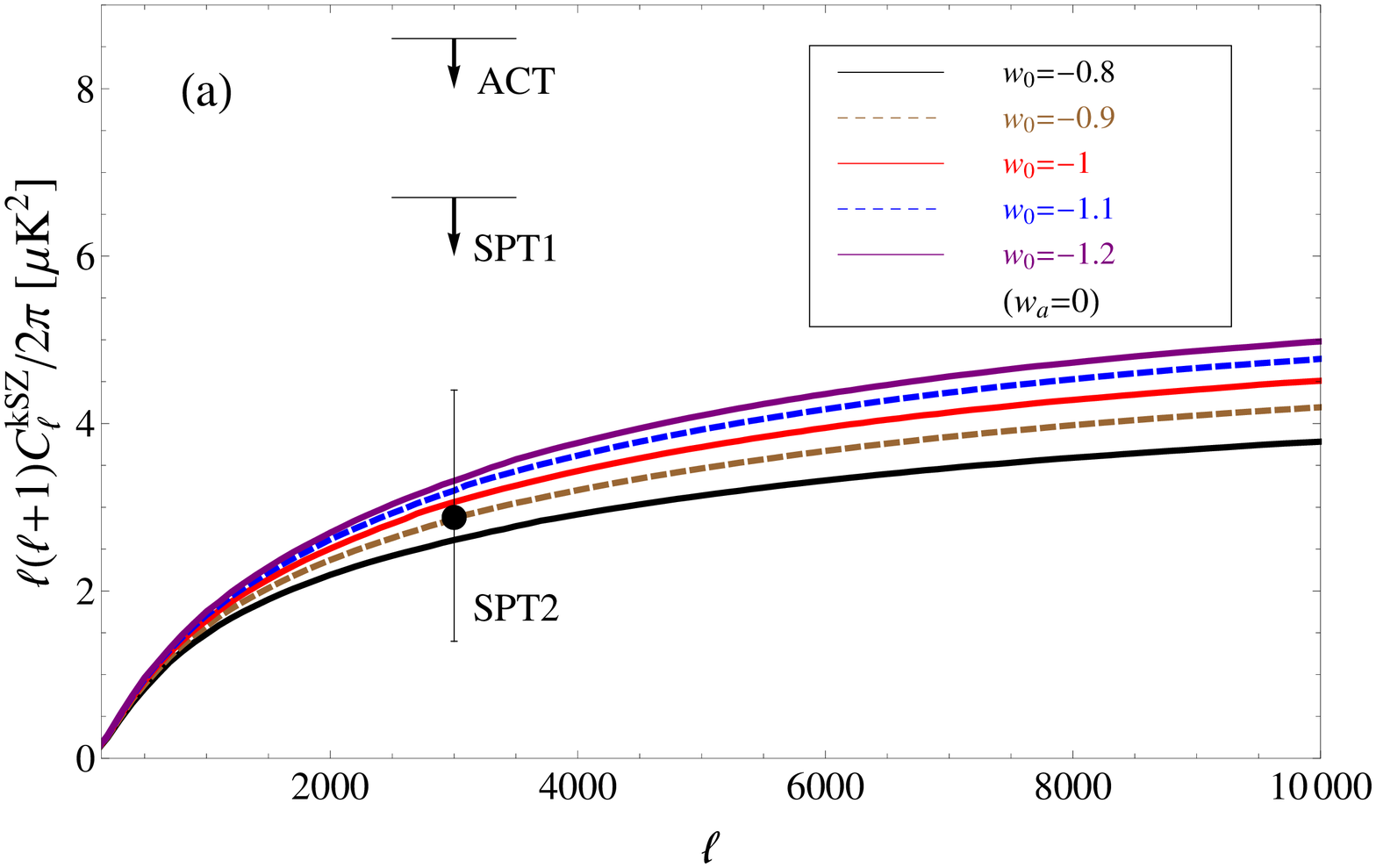}
\includegraphics[bb=0 0 862 450, width=4.1in]{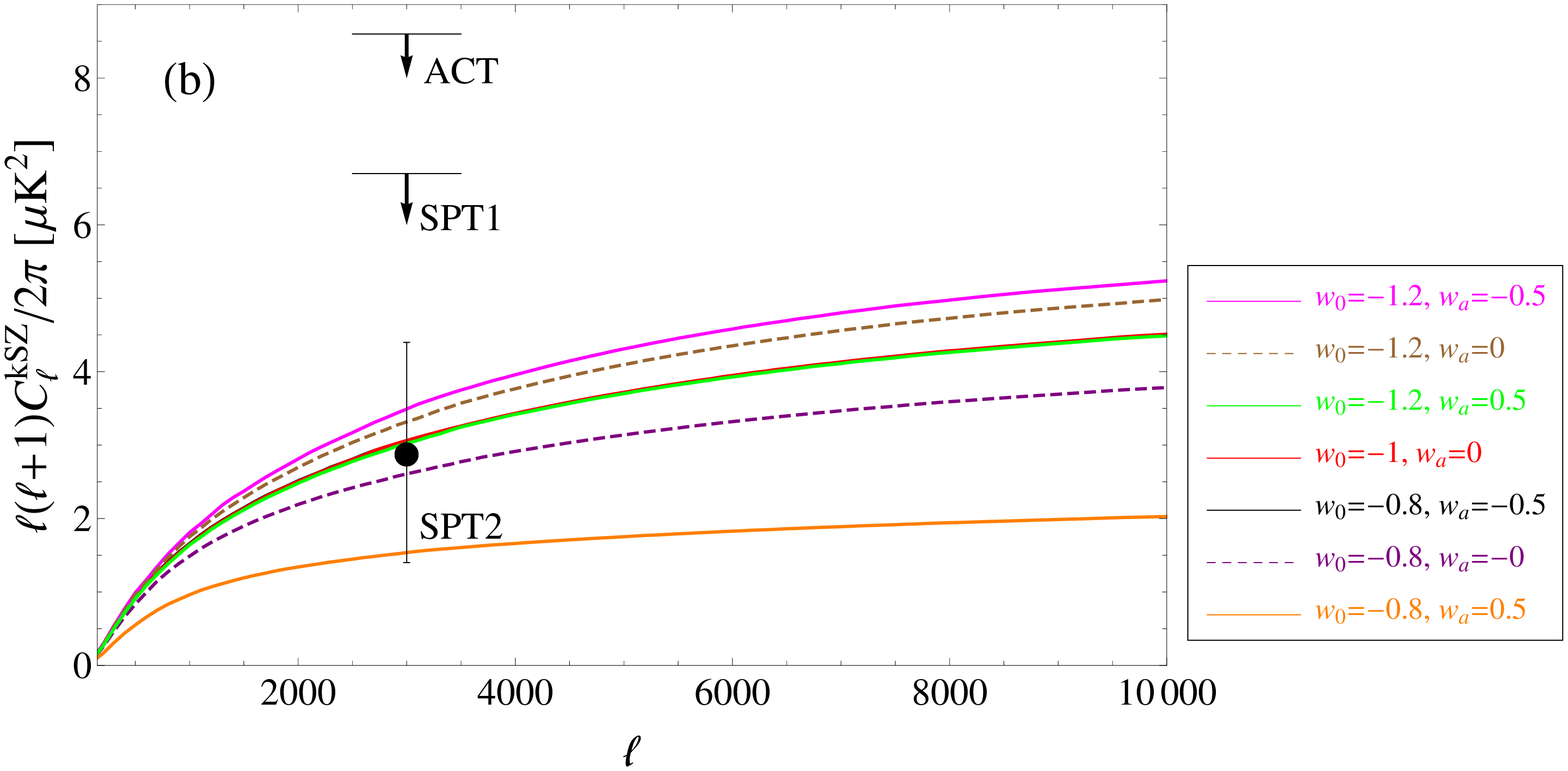}}
\caption{\textit{Panel(a)}: kSZ angular power spectrum
($D_{\ell}\equiv \ell(\ell+1)C^{\rm kSZ}_{\ell}/2\pi$) for five
dark energy models with constant EoS. \textit{Panel (b)}: Same as
panel~(a) but for time-varying dark energy models. The horizontal
bars with arrows show $D_{\ell=3000} \leq 8.6 \mu\textrm{K}^{2}$
($95\%$ confidence level) from Atacama Cosmology Telescope (ACT)
\cite{Sievers13} and $D_{\ell=3000} \leq 6.7 \mu\textrm{K}^{2}$
($95\%$ CL) from South Pole Telescope (SPT1) \cite{Reichardt12}.
The black data point  shows $D_{\ell=3000}=2.9 \pm 1.5
\mu\textrm{K}^{2}$ ($1\sigma$ CL) also from South Pole Telescope
(SPT2) while including bispectrum constraints \cite{Crawford13}.}
\label{fig:Dl2}
\end{figure*}

\begin{figure*}[tbp]
\centerline{
\includegraphics[bb=0 0 680 353, width=3.6in]{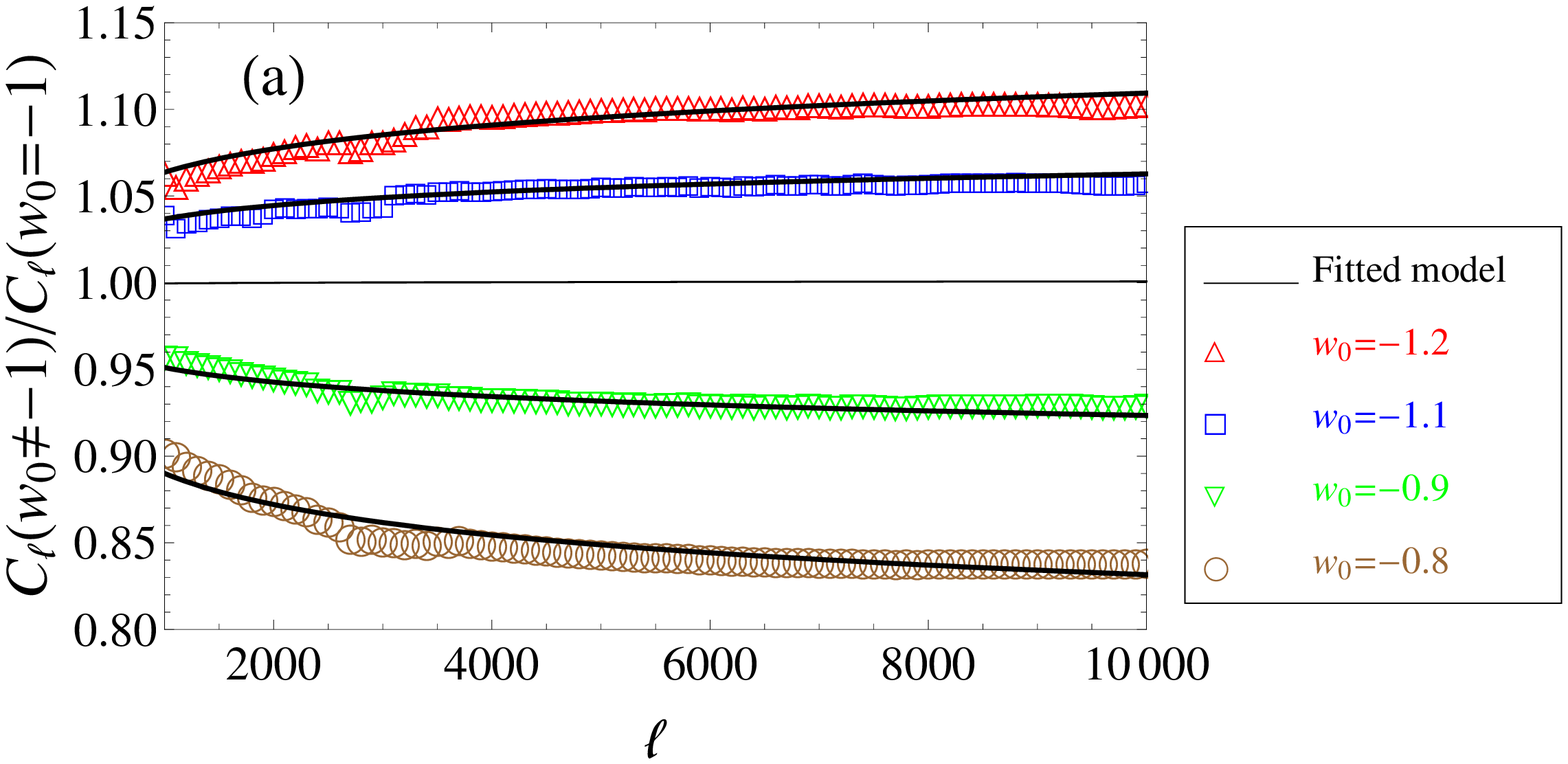}
\includegraphics[bb=0 0 748 368, width=3.8in]{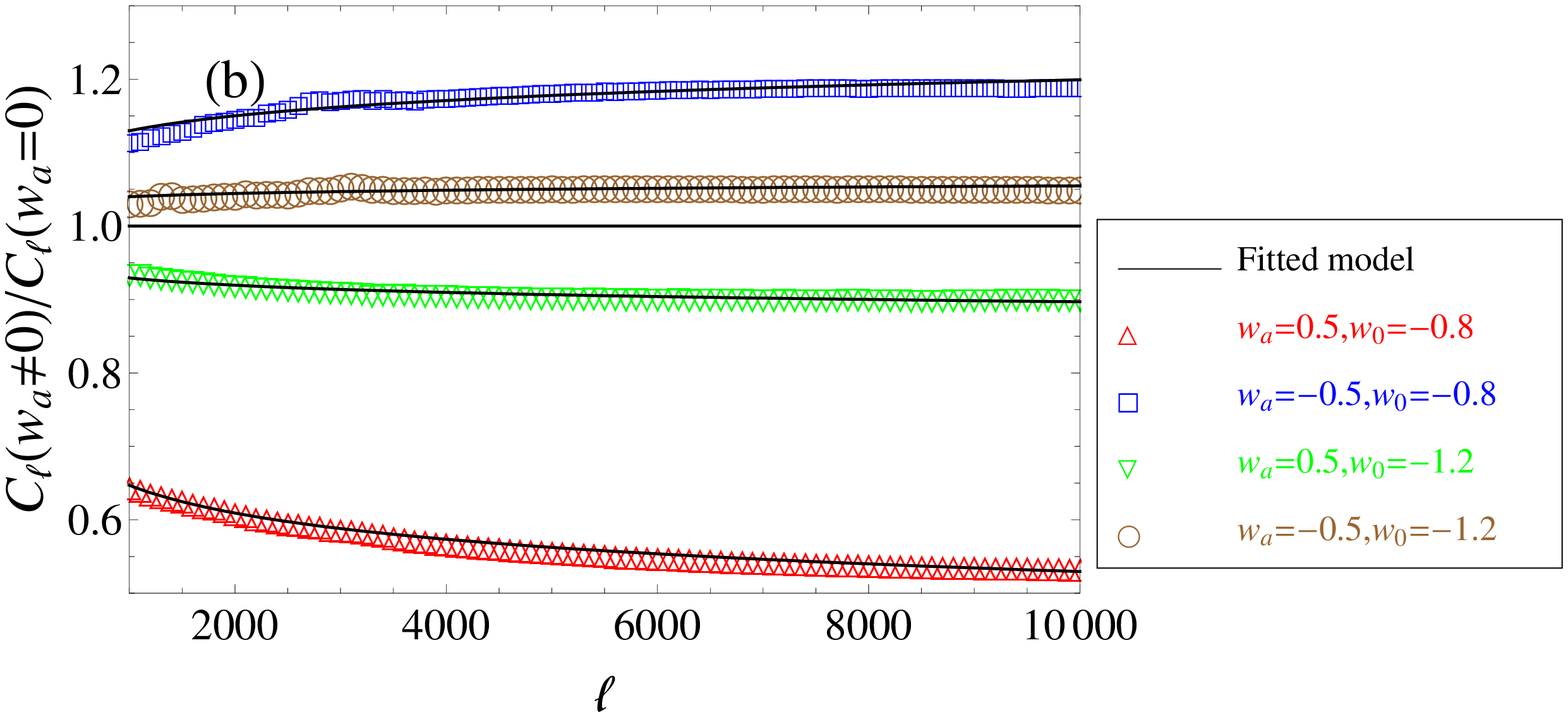}}
\caption{\textit{Panel(a)}: The ratio between $w$CDM and
$\Lambda$CDM $C_{\ell}$s (i.e. ratio between model with $w_{0}\neq
-1$ and $w_{0} = -1$) with the scaling law
(Eqs.~(\ref{eq:power-law}) and (\ref{eq:BC-1})) marked as black
lines. \textit{Panel (b)}: the ratio between $C_{\ell}$ with
$w_{a}\neq0$ and with $w_{a}=0$ (note that this is not ratio
between dynamical dark energy with $\Lambda$CDM model), the black
lines correspond to the scaling law (Eqs.~(\ref{eq:power-law2})
and (\ref{eq:BC-2})). The accuracy of the fit is within $1\%$ over
$\ell \simeq 3000$--$10000$.} \label{fig:scaling}
\end{figure*}

\subsection{3D power spectrum of curl component of momentum field}
\label{sec:kernel}

To calculate the 3D curl momentum power spectrum $\Delta_{\rm
b}(k)$ at different redshifts, we rewrite
Eq.~(\ref{eq:deltab2-NL}) as,
\begin{eqnarray}
\Delta^{2}_{\rm b}(k,z) &=& \frac{k^{3}}{2\pi^{2}}(\dot{a}f)^{2}
\int \frac{\der k' \der \mu}{(2\pi)^{2}} P^{\rm NL}_{\delta
\delta}(|\vec{k}-\vec{k}'|) \nonumber \\
& \times & P_{\delta \delta} \tilde{I}(k,k'),
\label{eq:deltab2-NL-modify1}
\end{eqnarray}
where
\begin{eqnarray}
\tilde{I}(k,k',\mu)=\frac{((k/k')^{2}-2\mu(k/k'))(1-\mu^{2})}{1+(k/k')^{2}-2\mu(k/k')},\label{eq:kernel-2}
\end{eqnarray}
is the reduced dimensionless kernel function. We plug in the
calculation of $f(z)$ and the linear and non-linear matter power
spectrum ($P_{\delta \delta}$ and $P^{\rm NL}_{\delta \delta}$)
into Eq.~(\ref{eq:deltab2-NL-modify1}), and integrate over the
cosine angle of separation $\mu=[-1,1]$ and $k'$, and then obtain
the 3D curl component of momentum power spectrum. We also
calculate the OV effect for comparison.

In Fig.~\ref{fig:deltab-lcdm}, we plot the power spectrum of
momentum field of the fiducial $\Lambda$CDM model at different
redshifts. It is obvious that more and more structures form as the
universe evolves, therefore the amplitude of curl momentum power
spectrum increases as redshift drops.
At high $z$, {\it e.g.}, $z=6$, the nonlinearity has less effect
on the kSZ $\Delta_{\rm b}$ on the concerning scales thus the
linear OV approach is a good approximation. However, as the
universe evolves, the rms of fluctuation exceeds unity on larger
and larger scales, so structures become non-linear on
comparatively larger scales. This makes the $z=0$ curl momentum
power spectrum significantly different from the OV power spectrum
on scales of $k>1 \mpch$.

Now we can compare $\Delta_{\rm b}(k)$ of $w$CDM
cosmology to that of the $\Lambda$CDM cosmology. In
Fig.~\ref{fig:deltab-w0}, we plot the fractional difference of
$w$CDM momentum power spectrum with the fiducial $\Lambda$CDM
model, at four different redshifts, which are chosen
as follows: $z=6$ as the onset of structure formation, $z=3$ as
the typical epoch of gravitational collapse, $z=1$ as the era when
dark energy becomes important, and $z=0$ represents the current
epoch. For each panel, we choose $w_{0}$ to be
[$-1.2$, $-1.1$, $-1$, $-0.9$, $-0.8$].

One can see that at $z=6$, there is little difference between
$w$CDM prediction and the $\Lambda$CDM prediction, since in both
scenarios the dark energy component is negligible. The dark
energy effect kicks in at $z=1$, 
making the
fractional difference 
reach $5\%$ at this time. At even later
time, this difference become more significant, and at present time
this different is $\sim10\%$.

We show the fractional difference between CPL dark energy model
and $\Lambda$CDM model in Fig.~\ref{fig:deltab-w0-wa}. One can see
that the more negative $w_{0}$ or $w_a$ is, the higher the
amplitude of curl momentum field is, and vice versa. This is
natural since $\Delta_{\rm b}(k)$ increases as the matter power.

\subsection{The total signal}
\label{sec:total}

\begin{table*}
\begin{centering}
\begin{tabular}{@{}|l|l|l|l|l|l|}\hline
$D_{\ell}(\mu {\rm K}^{2})$& $w_{0}=-0.8$ & $w_{0}=-0.9$ &
$w_{0}=-1$ & $w_{0}=-1.1$ & $w_{0}=-1.2$ \\ \hline $ \ell=  1000$
& $ 1.49$ & $ 1.58$ & $ 1.65$ & $ 1.71$ & $ 1.76$ \\ \hline $
\ell=  2000$ & $ 2.19$ & $ 2.37$ & $ 2.51$ & $ 2.62$ & $ 2.70$ \\
\hline $ \ell= 3000$ & $ 2.61$ & $ 2.86$ & $ 3.06$ & $ 3.20$ & $
3.32$ \\ \hline $ \ell=  4000$ & $ 2.92$ & $ 3.20$ & $ 3.43$ & $
3.62$ & $ 3.77$
\\ \hline $ \ell=  5000$ & $ 3.14$ & $ 3.46$ & $ 3.72$ & $ 3.93$ &
$ 4.10$ \\ \hline $ \ell=  6000$ & $ 3.32$ & $ 3.67$ & $ 3.95$ & $
4.17$ & $ 4.35$ \\ \hline $ \ell=  7000$ & $ 3.47$ & $ 3.84$ & $
4.13$ & $ 4.37$ & $ 4.56$ \\ \hline $ \ell=  8000$ & $ 3.59$ & $
3.98$ & $ 4.28$ & $ 4.53$ & $ 4.73$ \\ \hline $ \ell=  9000$ & $
3.69$ & $ 4.10$ & $ 4.40$ & $ 4.66$ & $ 4.87$ \\ \hline $ \ell=
10000$ & $ 3.78$ & $ 4.20$ & $ 4.51$ & $ 4.77$ & $ 4.98$ \\ \hline
\end{tabular}%
\caption{The values of kSZ power spectrum
$D_{\ell}=\ell(\ell+1)C_{\ell}/2\pi\, [\mu {\rm K}^{2}]$ of $10$
multiples for five constant $w$ models.} \label{tab:w0-compare}
\end{centering}
\end{table*}

\begin{table*}
\begin{centering}
\begin{tabular}{@{}|l|l|l|l|l|l|l|}\hline
$D_{\ell}(\mu {\rm K}^{2})$& $w_{a}=0.5$ & $w_{a}=0$ &
$w_{a}=-0.5$ &
$w_{a}=0.5$ & $w_{a}=0$ & $w_{a}=-0.5$ \\
& $w_{0}=-0.8$ & $w_{0}=-0.8$ & $w_{0}=-0.8$ & $w_{0}=-1.2$ &
$w_{0}=-1.2$ & $w_{0}=-1.2$ \\ \hline $ \ell=  1000$ & $ 0.96$ & $
1.49$ & $ 1.66$ & $ 1.64$ & $ 1.76$ & $ 1.81$ \\ \hline $ \ell=
2000$ & $ 1.34$ & $ 2.19$ & $ 2.52$ & $ 2.48$ & $ 2.70$ & $ 2.81$
\\ \hline $ \ell=  3000$ & $ 1.54$ & $ 2.61$ & $ 3.06$ & $ 3.02$ &
$ 3.32$ & $ 3.49$ \\ \hline $ \ell=  4000$ & $ 1.66$ & $ 2.92$ & $
3.43$ & $ 3.42$ & $ 3.77$ & $ 3.96$ \\ \hline $ \ell=  5000$ & $
1.75$ & $ 3.14$ & $ 3.71$ & $ 3.70$ & $ 4.10$ & $ 4.31$ \\ \hline
$ \ell=  6000$ & $ 1.83$ & $ 3.32$ & $ 3.94$ & $ 3.93$ & $ 4.35$ &
$ 4.58$ \\ \hline $ \ell=  7000$ & $ 1.89$ & $ 3.47$ & $ 4.12$ & $
4.11$ & $ 4.56$ & $ 4.80$ \\ \hline $ \ell=  8000$ & $ 1.94$ & $
3.59$ & $ 4.27$ & $ 4.26$ & $ 4.73$ & $ 4.98$ \\ \hline $ \ell=
9000$ & $ 1.99$ & $ 3.69$ & $ 4.40$ & $ 4.38$ & $ 4.87$ & $ 5.12$
\\ \hline $ \ell= 10000$ & $ 2.03$ & $ 3.78$ & $ 4.50$ & $ 4.49$ &
$ 4.98$ & $ 5.24$ \\ \hline
\end{tabular}%
\caption{The values of kSZ power spectrum
$D_{\ell}=\ell(\ell+1)C_{\ell}/2\pi\, [\mu{\rm K}^{2}]$ of $10$
multiples for varying $w_{a}$ models.} \label{tab:wa-compare}
\end{centering}
\end{table*}

Now we put together the factors of structure growth, comoving
distance, and power spectrum of curl momentum field to analyze how
dark energy affects the kSZ angular power spectrum.

Note that Eq.~(\ref{eq:ksz-cls}) is an integral up to $z_{\rm
rei}=10$, so it is a projected effect of the velocity field along
line of sight. Therefore, we need to count for all the observable
modes of fluctuations at different redshifts. By calculating
$\der^{2}C_{\ell}/\der z \der \ln k$, Ref.~\cite{Shaw12} shows (in
their Fig.~1) that, $75\%$ of the full kSZ power comes from
redshifts in the range of $[0,7]$ and $k$-mode in the range of
$0.2-7.0 \mpch$ at $\ell=3000$.
This $\{k,z\}$ range is ideal to probe for the amplitude and even
the time evolution of the dark energy EoS, thus the kSZ
measurement can potentially facilitate a novel test of dark
energy. Note that although in the following we plot the
$C_{\ell}$s up to $\ell \simeq 10000$, most of the constraining
power related to cosmology comes from $\ell \simeq 3000$.

In Fig.~\ref{fig:Dl2} (a), we plot the kSZ angular power spectrum
$\ell(\ell+1)C_{\ell}/2\pi$ as a function of the multipole $\ell$.
We show the result for the various dark energy models with a
constant EoS from $-0.8$ to $-1.2$. One can see that, since a more
negative $w_{0}$ makes the comoving distance $x(z)$, growth
function $f(z)$ and amplitude of curl momentum field $\Delta_{\rm
b}$(k) coherently larger, the cumulative integral will eventually
enhance the total signal $C_{\ell}$ significantly, and vice versa.
On scales of $\ell \sim 3000$, $C_{\ell}(w=-0.8)$ is smaller than
the $\Lambda$CDM value by a factor of $14.7\%$, while
$C_{\ell}(w=-1.2)$ is larger than the $\Lambda$CDM value by a
factor of $8.5\%$. So the total variation of signal given the
allowed parameter space by \textit{WMAP} observations
\cite{Hinshaw12} can reach nearly $23\%$ on scales of $\ell=3000$.
On even smaller scales (larger $\ell$'s), the difference can be
even more significant. We list the values of $C_{\ell}$'s of $10$
multiples separated by $\Delta \ell=1000$ in
Table~\ref{tab:w0-compare}. This is the most prominent effect of
dark energy on kSZ power spectrum.

In addition, in Fig.~\ref{fig:Dl2}~(b), we plot the kSZ power
spectrum for dark energy models with a non-zero $w_a$, namely,
$w_{a}$ from $-0.5$ to $0.5$. Note that this range of $w_{a}$
value is allowed by the joint constrained from {\it
WMAP}9+SPT+ACT+BAO+$H_{0}$ \cite{Lambdaweb}. We can see that with
$w_{0}=-0.8$ or $w_{0}=-1.2$, if $w_{a}>0$ the dark energy kicks
in earlier than $w_{a}=0$, so the structure growth will be
suppressed and vice versa. Quantitatively, the change in $w_{a}$
by $\pm 0.5$ results in a change in the kSZ signal by a factor of
$50$ to $60\%$ on scales of $\ell=3000$, which is a significant
effect manifesting the properties of dark energy. We list the $10$
values of $C_{\ell}$'s for CPL dark energy model in
Table~\ref{tab:wa-compare}.

To use the kSZ measurements to constrain dark energy EoS, one
needs to calculate the kSZ power spectra for a large numbers of
cosmological models for the Markov Chain Monte Carlo (MCMC)
process. This is computationally expensive so it is useful to
develop accurate fitting formula for the practicality.


To understand the feature of kSZ spectrum, we plot the ratio of the spectrum between the
constant-$w$ model and the $\Lambda$CDM in
Fig.~\ref{fig:scaling} (a) where dots in different colours represent
different values of $w$. We can see that the trend of
$C_{\ell}(w\neq-1)/C_{\ell}(w=-1)$ close to a power law shape, so
we model the function as
\begin{eqnarray}
\frac{C_{\ell}(w\neq-1)}{C_{\ell}(w=-1)}=B\left(\frac{\ell}{1000}
\right)^{C}, \label{eq:power-law}
\end{eqnarray}
where the amplitude $B$ and the power index $C$ are to be
determined. We first output the left-hand-side of
Eq.~(\ref{eq:power-law}) for each $\ell$, and then by assuming a
form of $B$ and $C$ as $a_{1}+a_{2}\exp(w)+a_{3} w$, we fit these
parameters with the data $C_{\ell}(w \neq -1)/C_{\ell}(w=-1)$. We
find that the following function can very well approximate the
function,
\begin{eqnarray}
B &=& 2.84-3.08\exp(w)+0.70 w, \nonumber \\
C &=& 0.49-0.83 \exp(w)+0.19 w .\label{eq:BC-1}
\end{eqnarray}
In Fig.~\ref{fig:scaling} (a), we compare the exact numerical results of
$C_{\ell}(w)/C_{\ell}(\Lambda{\rm CDM})$ as colour dots with the above
fitting formula (Eqs.~(\ref{eq:power-law}) and (\ref{eq:BC-1}))
as black solid lines. One can find an excellent agreement between
the two.

Furthermore, we investigate the empirical relation between
$w_{0}$-$w_{a}$ dark energy kSZ signal with fiducial $\Lambda$CDM
model. In Fig.~\ref{fig:scaling} (b), we plot the ratio between the
kSZ power spectrum with $w_{a}\neq 0$ and the one with $w_{a}=0$.
The colour scheme represents different values of ($w_{0}$,
$w_{a}$). One can see that this ratio function is also close to a
power law form, we therefore parameterize it as
\begin{eqnarray}
\frac{C_{\ell}(w_{a}\neq0)}{C_{\ell}(w_{a}=0)}=B'\left(\frac{\ell}{1000}
\right)^{C'}. \label{eq:power-law2}
\end{eqnarray}
Then we find that if allowing $B'$ and $C'$ related to a parameter
$\tilde{x}=w_{a}/w_{0}$, then the ratio function can be well
approximated by (by using the same fitting method as described
above)
\begin{eqnarray}
B'(\tilde{x}) &=& 22.43+21.35 \tilde{x}+10.78 \tilde{x}^{2} \nonumber \\
 &+& 4.84 \tilde{x}^{3}-21.43\exp(\tilde{x}) , \nonumber \\
C'(\tilde{x}) &=& 5.58(1-\exp(\tilde{x})) +5.55 \tilde{x}
\nonumber \\
&+& 2.8 \tilde{x}^{2}+1.25 \tilde{x}^{3} .\label{eq:BC-2}
\end{eqnarray}
In Fig.~\ref{fig:scaling} (b), we compare the numerical values of
the ratio function by colour dots and its empirical relation
(\ref{eq:power-law2}) and (\ref{eq:BC-2}) by black solid lines. We
again find an excellent agreement between the two. Therefore, our
fitting formulae (Eqs.~(\ref{eq:power-law})--(\ref{eq:BC-2})) can
be used for fast calculation of models with $w_{0} \neq -1$ and
$w_{a}\neq 0$. Here, we remind the reader that the scaling
relation between $C^{\rm kSZ}_{\ell}$ and other cosmological
parameters ({\it e.g.} $\Omega_{\rm b}$, $\sigma_{8}$, $z_{\rm
rei}$, and $\tau$) is investigated in \cite{Shaw12}, so can also
be used in fast numerical computation.

\subsection{Observational constraints}
\label{sec:observe}

We now discuss what current and future observational constraints
can be obtained on the kSZ power spectrum and its prospective to
constrain dark energy. In \cite{Sievers13}, by using 148 GHz and
218 GHz Atacama Cosmology Telescope (ACT) data and fitting the
template with contribution from thermal and kinetic SZ effects,
infrared sources and radio sources, the $95\%$ level upper limit
is found to be $8.6 \mu\textrm{K}^{2}$. In \cite{Reichardt12}, the
constraint on $D_{\ell=3000}$ is obtained by combining $95$, $150$
and $220$ GHz channel data of SPT. By fitting the template of
thermal SZ with the kinetic SZ signal, it is found that $D^{\rm
kSZ}_{\ell=3000}<2.8 \mu {\rm K}^{2}$ at $95\%$ CL. In addition,
if considering the the correlation between thermal SZ effect with
cosmic infrared background, this upper limit is loosed to $D^{\rm
kSZ}_{\ell=3000}<6.7 \mu {\rm K}^{2}$ \cite{Reichardt12} at $95\%$
CL. Furthermore, by incorporating the bispectrum data from the
same three channels of SPT, Ref.~\cite{Crawford13} finds that the
derived constraints on kSZ amplitude at $\ell=3000$ is $D^{\rm
kSZ}_{\ell=3000}=2.9 \pm 1.5 \mu {\rm K}^{2}$ at $1\sigma$
confidence level (CL.), and $<5.5 \mu {\rm K}^{2}$ at $95\%$ CL.
We place these upper limits and data point in the two panels of
Fig.~\ref{fig:Dl2}.

By comparing the constraints from \cite{Sievers13},
\cite{Reichardt12} and \cite{Crawford13}, we can see that although
the constraints are not very strong at current situation, the SPT
constraint with bispectrum (black data point on
Fig.~\ref{fig:Dl2}b) already tend to rule out the model with
($w_{0}=-0.8$, $w_{a}=0.5$). In addition, the trend of tightening
constraints of kSZ signal is quite obvious given many of the
ongoing CMB surveys. In the future, if we can place both upper
and lower limit on kSZ power
spectrum, 
it can be used as a powerful tool to constrain EoS of dark energy.
In reality, \textit{Herschel} data can be used to separate the
infrared and radio sources in the foreground, and thus improve the
constraints on kSZ signal.

\subsection{Relation to patchy reionization}
\label{sec:patchy}

What we modeled above is the homogeneous kSZ signal which comes
from the era after reionization $z\lesssim 10$. The total signal
of kSZ consists of both homogeneous kSZ signal and the patchy
reionization signal with most of its contribution from reionization
era. The magnitude of the kSZ power from the second component,
i.e. patchy reionization, is strongly related to the process of
reionization \cite{Zahn05,McQuinn05}, which detail is relatively
unknown. For instance, it is unclear whether the reionization is
an instantaneous reionization, or two-step reionization, or a
double reionization \cite{Zahn12,Pritchard10}. In addition, it is not
clear how much contribution of the total kSZ signal from the
patchy reionization era. For example, if reionization started at
$z=14$ and ended at $z=6$, then it can generate roughly $3 \mu
{\rm K}^{2}$ of patchy kSZ power (at $\ell \simeq 3000$), while
the range $z=[8,12]$ would generate $1.5 \mu {\rm K}^{2}$
\cite{McQuinn05,Shaw12}. Therefore, in order to derive the patchy
component of total kSZ signal, it is very important to have a good
theoretical modeling of the homogeneous kSZ contribution as was
laid out in this paper.

\section{Conclusion}
\label{sec:conclude}

The nature of dark energy
is a mystery in modern cosmology, and its property 
is characterized by its equation of
state (EoS) parameter. Current CMB space-mission such as
\textit{WMAP} and \textit{Planck}, ground-based CMB experiments
such as ACT and SPT, as well as baryon acoustic oscillation
experiments from SDSS can set up tight constraints on $w$
parameter if assuming that $w$ is a constant. However, if allowing
$w$ to vary, such as $w(a)=w_{0}+w_{a}(1-a)$ (the CPL
parametrization), the constraints become weaker while a large
region of parameter space is allowed.

In this paper we have calculated the kinetic Sunyaev-Zel'dovich
signal for general dark energy models with both the constant-$w$
case, and the CPL parametrization (time-varying $w$) case. We
first review the calculation of the kSZ signal for the
$\Lambda$CDM model, and extend the analysis for the general dark
energy model.


We calculate the curl momentum power spectrum $\Delta_{\rm b}(k)$
at different redshifts, and find that dark energy can affect the
amplitude and shape of the gravitational clustering at redshifts
$0-3$. Finally, we integrate the curl momentum field from redshift
$0$ till the reionization redshift $z_{\rm rei}=10$, and find that
if, for example, $w_{0}=-0.8$ the total signal of kSZ can be
suppressed by a factor of $\sim 14.7\%$ on scales of $\ell=3000$,
while $w_{0}=-1.2$ the total signal of kSZ can be enhanced by a
factor of $\sim 8.5\%$ on the same scales. We then vary the
parameter $w_{a}$ and find that this parameter is more sensitive
to the amplitude and shape of the kSZ signal, and in the range of
$w_{a}=\pm 0.5$ ($1\sigma$ constrained parameters space by {\it
WMAP}9+ACT+SPT+BAO+$H_{0}$), the $w_{a}$ can alter the amplitude
of kSZ signal by nearly $60 \%$. Therefore, if kSZ signal can be
precisely measured, it can be a sensitive test of dark energy.

Finally, in order to fast calculate the kSZ signal in a general
dark energy model with a constant $w$ or a time-varying $w$, we
model an empirical relation which can precisely recover the values
of kSZ power spectrum from numerical calculation. Our fitting
formulae (Eqs.~(\ref{eq:power-law})-(\ref{eq:BC-2})) work very
precisely in a large region of parameter space ($w_{0}$, $w_{a}$)
and therefore can be useful in the fast computation of $C^{\rm
kSZ}_{\ell}$.

\section{Acknowledgement}
We thank the helpful discussion with Douglas Rudd, Laurie Shaw and
Pengjie Zhang. Y.Z.M. is supported by a CITA National Fellowship and
the Natural Science and Engineering Research Council of Canada.
G.B.Z. is supported by the {\it 1000 Young Talents} Fellowship in China, 
by the 973 Program
grant No. 2013CB837900, NSFC grant No. 11261140641,
and CAS grant No. KJZD-EW-T01, and by the Strategic Priority Research Program 
``The Emergence of Cosmological Structures'' of the Chinese Academy of Sciences, Grant No. XDB09000000.

\appendix
\section{Derivation of $\chi$ and $\mu_{\rm e}$}
\label{sec:chi-mu-derive} In Section~\ref{sec:kSZ_effect}, we
define $\chi$ as the fraction of the total number of electrons
that are ionized. We assume that at $z<z_{\rm rei}$ the hydrogen
is completely ionized, and the number of helium electrons ionized
is $N_{\rm H_{e}}$, so $N_{\rm H_{e}}$ can take $0$, $1$ and $2$
for neutral, singly and fully ionized helium respectively. In our
fiducial model we assume $N_{\rm H_{e}}=0$ at all redshifts. Thus
$\chi$ is the ratio between ionized and total number of electrons,
i.e.
\begin{eqnarray}
\chi &=& \frac{\overline{n}_{\rm e,i}}{n_{\rm e}} \nonumber \\
&=& \frac{n_{\rm H}+ n_{\rm H_{e}}\cdot N_{\rm H_{e}}}{n_{\rm H}+2
n_{\rm H_{e}}}. \label{eq:chi-def}
\end{eqnarray}
The helium number density is
\begin{eqnarray}
n_{\rm H_{e}}=\frac{Y_{\rm p}}{4 X_{\rm p}}n_{\rm H},
\label{eq:helium-n}
\end{eqnarray}
where $Y_{\rm p}=0.24$ and $X_{\rm p}=1-Y_{\rm p}$ is the
primordial helium and hydrogen abundance. Therefore substituting
Eq.~(\ref{eq:helium-n}) into Eq.~(\ref{eq:chi-def}), we obtain
\begin{eqnarray}
\chi =\frac{1-Y_{\rm p}(1-N_{\rm H_{e}}/4)}{1-Y_{\rm p}/2}.
\label{eq:chi-def2}
\end{eqnarray}

We now calculate the gas density as
\begin{eqnarray}
\rho_{\rm g}=m_{\rm p}n_{\rm H}+ m_{\rm H_{e}} n_{\rm H_{e}},
\label{eq:rho-g1}
\end{eqnarray}
since $m_{\rm H_{e}} \simeq 4 m_{\rm p}$, and by using
Eq.~(\ref{eq:helium-n}), we obtain
\begin{eqnarray}
n_{\rm H}=\frac{\rho_{\rm g}}{\left(1+\frac{Y_{\rm p}}{X_{\rm p}}
\right)m_{\rm H}} . \label{eq:n-H}
\end{eqnarray}
Since the total electron density is $n_{\rm e}=n_{\rm H}+2n_{\rm
H_{e}}$, by using Eq.~(\ref{eq:helium-n}), we obtain
\begin{eqnarray}
n_{\rm e}=\frac{\rho_{\rm g}}{m_{\rm p}\mu_{\rm e}}, \text{
}\mu_{\rm e}=\frac{1+ Y_{\rm p}/X_{\rm p}}{1+ Y_{\rm p}/(2X_{\rm
p})}=1.14, \label{eq:mue}
\end{eqnarray}
where $\mu_{\rm e}$ is called mean electron weight. Then combining
Eqs.~(\ref{eq:chi-def}), (\ref{eq:chi-def2}) and (\ref{eq:mue}),
we obtain
\begin{equation}
\overline{n}_{\rm e,i}=\frac{\chi \rho_{g}}{m_{\rm p}\mu_{\rm e}}.
\end{equation}

\end{document}